\documentclass[conf]{new-aiaa}
\usepackage[utf8]{inputenc}
\usepackage[margin=1in]{geometry}
\usepackage{graphicx}
\usepackage{amsmath}
\usepackage{mathtools}
\usepackage[version=4]{mhchem}
\usepackage{siunitx}
\usepackage{longtable,tabularx}
\usepackage{soul,xcolor}
\usepackage{float}
\usepackage{caption}
\usepackage{subcaption}
\usepackage{gensymb}
\usepackage{nomencl}
\usepackage[skip=10pt plus1pt, indent=40pt]{parskip}
\usepackage{natbib}
\usepackage{placeins}
\usepackage[capitalize]{cleveref}
\usepackage{xspace} 

\newcommand{\rev}[1]{{\color{black}{#1}}}
\newcommand{\DGSEM}{\textsc{DGSEM}\xspace}
\newcommand{\FVSE}{\textsc{FVSE}\xspace}
\newcommand{\EV}{\textsc{EV}\xspace}
\newcommand{\Hybrid}{{Hybrid}\xspace}
\newcommand{\LLF}{{LLF}\xspace}

\def\FirstAuthor{Zachary Pyle\footnote{Doctoral Student, Department of Aerospace Engineering, zpyle8687@sdsu.edu, AIAA Student Member}}
\def\SecondAuthor{Gustaaf B. Jacobs\footnote{Professor, Department of Aerospace Engineering, gjacobs@sdsu.edu, Associate Fellow, Corresponding Author}}

\def\FirstAffil{San Diego State University, San Diego, CA, 92182}

\author{\FirstAuthor and \SecondAuthor}
\affil{\FirstAffil}

\title{Robust Spectral Solver for High-Fidelity Investigations of Aerospike Nozzle Flow Dynamics}
\date{\today}

\begin{document}

\maketitle

\begin{abstract}
A spectral element solver is developed for the high-fidelity simulation of the unsteady flow over an aerospike nozzle. 
The Navier-Stokes solver is a kinetic-energy-preserving, discontinuous Galerkin spectral element method (\DGSEM) combined with a hybridization of an entropy viscosity (\EV) and a finite-volume subcell element (\FVSE) shock-capturing scheme. 
The diffusive FVSE method is locally called only at locations where the \EV method cannot sufficiently smooth the sharp solution gradients that suddenly appear in the supersonic, vortex-dominated jet generated by the aerospike nozzle. 
Two-dimensional tests of a perfectly expanded and an underexpanded nozzle flow demonstrate that the method is high-order accurate and captures unsteady flow phenomena at supersonic and hypersonic conditions.
A resolved three-dimensional simulation at a Reynolds number of $95{,}000$ shows that the solver implicitly models turbulent dissipation at the subgrid scales.
To the authors' knowledge, these simulations represent the first \DGSEM computations of the resolved, unsteady flow over an aerospike nozzle.
\end{abstract}

\section{Introduction}

The revival of long-term space initiatives in both the public and private sectors is driving rapid advances in launch vehicle technology.
Additively manufactured injector channels reduce pressure loss and increase propulsive efficiency.
First-stage boosters autonomously return to recovery platforms to reduce launch costs and turnaround time. 
Full-flow, staged-combustion cycle engines, powered by turbopumps and preburners, produce higher specific impulses ($I_{sp}$) than pressure-fed or gas-generator cycle engines.
In stark contrast, the modern bell nozzle retains the same fundamental limitations as the original V-2 rocket.
The contour of the bell nozzle constrains the expansion of the exhaust plume, yielding a maximum $I_{sp}$ at a specific design altitude.
At lower altitudes, shocks form within the overexpanded jet, which incur losses in total pressure and reduce propulsive efficiency.

A promising successor to the bell nozzle is the altitude-compensating aerospike nozzle.
These nozzles expand the exhaust plume in response to changing ambient pressure and prevent overexpansion at low altitudes.
The altitude compensation improves the specific impulse through the flight envelope \cite{Ito2002_JPP, MarshallFactSheet, Mueller1971_UniversityOfND, Hagemann1998_JPP}.
The thrust-to-weight ratio of aerospike nozzles can be improved by truncating the spike tip \cite{Ito2002_JPP, Verma2009_JPP, Verma2011_JPP}.

\begin{figure}
    \centering
    \begin{minipage}{1.0\textwidth}
        \begin{subfigure}[b]{0.425\textwidth}
            \includegraphics[width=1.0\linewidth]{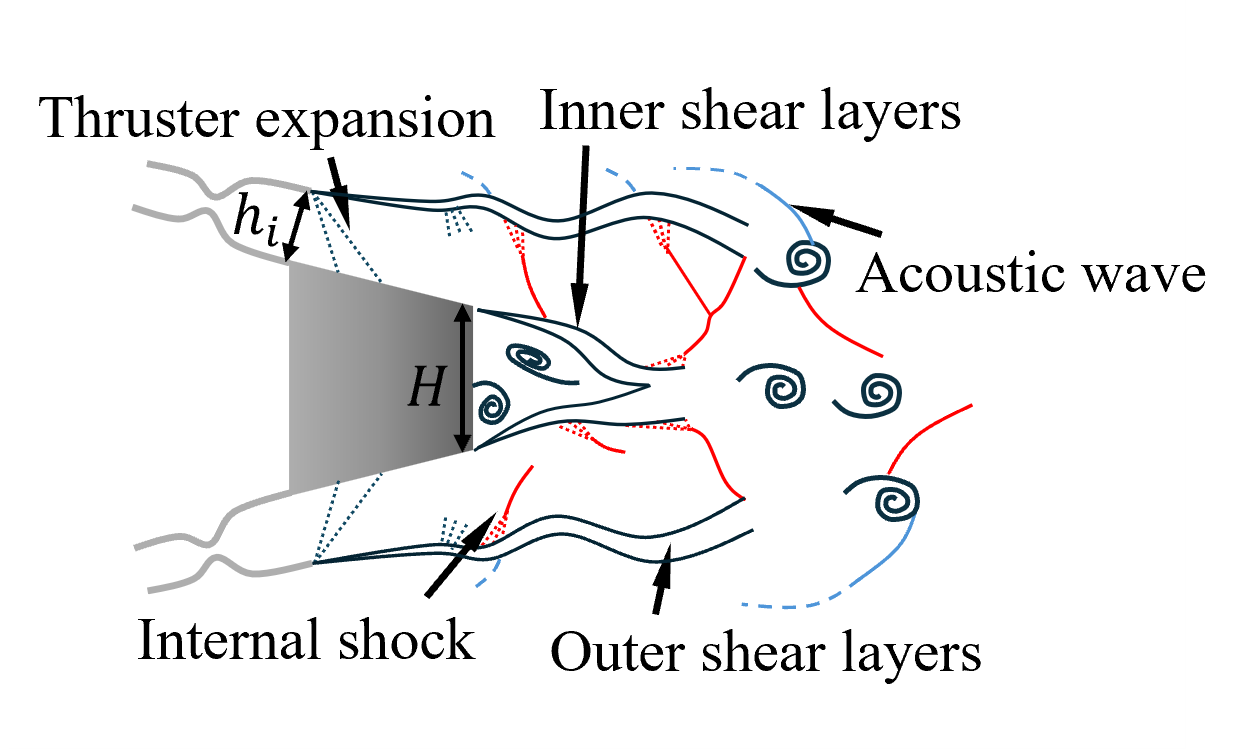}
            \caption{ }
            \label{fig:OpenWakeSchematic}
        \end{subfigure}
        \hfill
        \begin{subfigure}[b]{0.425\textwidth}
            \includegraphics[width=1.0\linewidth]{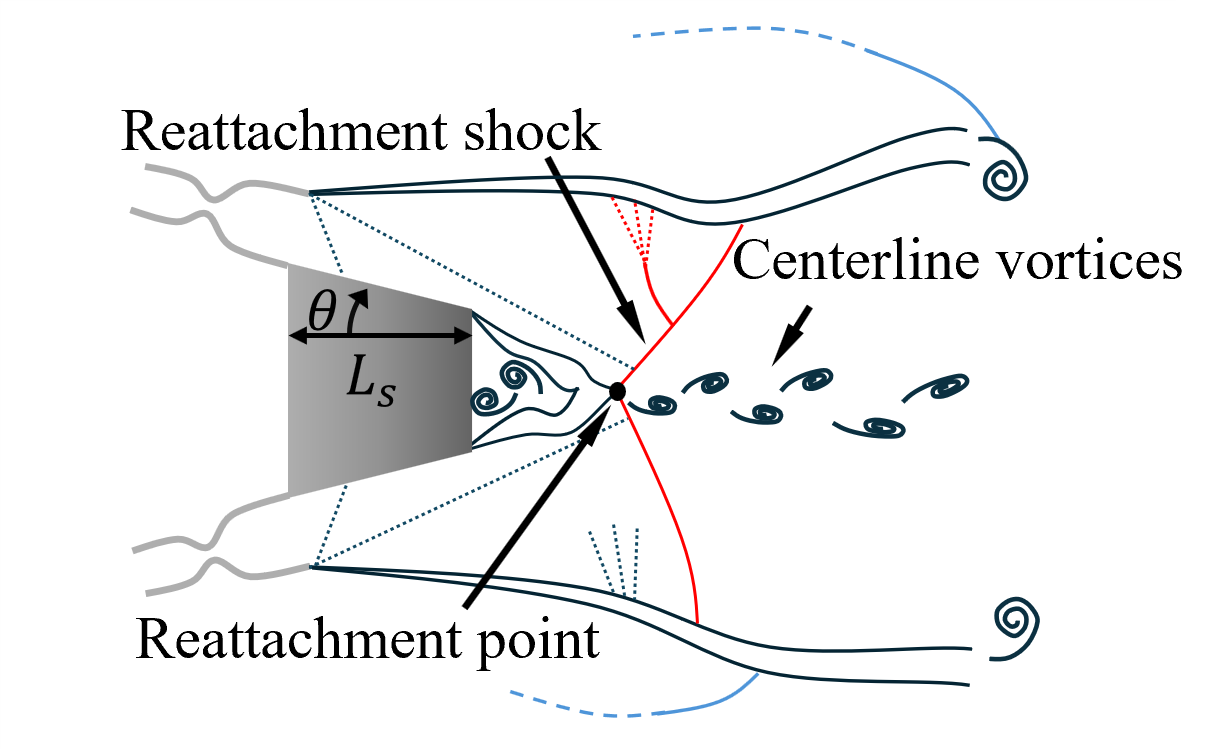}
            \caption{ }
            \label{fig:ClosedWakeSchematic}
        \end{subfigure}
    \end{minipage}
    \caption{Schematic of the (a) open wake and (b) closed wake.
    }
    \label{fig:ASSchematic}
\end{figure}

One key concern with the truncated aerospike is the unsteady flow it generates  and the related impact on flight stability. 
The unsteadiness originates from the interaction of the wall-bounded supersonic jets, which expand over the aerospike nozzle's ramp, at the so-called reattachment point (as illustrated in \cref{fig:ASSchematic}). 
At this location, a reattachment shock forms that interacts with instabilities along the shear line that also originate from the reattachment point.
A truncation of  the aerospike allows a base recirculation region to form in the near-wake region. 
Depending on flight conditions and nozzle design, this can lead to  an unsteady far-wake (\cref{fig:ASSchematic}). 
Schlieren images of  experiments on a linear aerospike nozzle flow conducted by  Mueller et al. \cite{Mueller1971_UniversityOfND}, Mueller et al. \cite{Mueller1972_UniversityOfND}, and Mueller and Sule \cite{Mueller1973_ASME} show that,  depending on the ratio of the nozzle stagnation pressure to the ambient pressure, $p_o/p_a$, a  vortex-dominated "open wake" or a quasi-steady "closed wake" is observed.
The wake is found to be open at lower nozzle pressure ratios, e.g., at flight conditions in the lower atmosphere and at higher ambient pressures, and a closed wake forms at higher nozzle pressure ratios, e.g., in the upper atmosphere. 
In their experiments, Verma et al. \cite{Verma2009_JPP} show distinct frequencies in the unsteady drag force on the aerospike nozzle if the wake is open.
They show that a  closed wake produces a broadband signal and that the power of the primary frequencies is reduced by an order of magnitude compared to the open wake. 
Although these studies report that the open wake is unsteady, they do not discuss the causality or mechanics of the unsteady flow.

\rev{

From inviscid flow analysis, Hagemann and collaborators   \cite{Hagemann1998_AIAAConf, Hagemann1998_JPP, Hagemann2001_AIAAConf}
posit that wake closure occurs if the trailing Mach wave of the centered, Prandtl-Meyer expansion fan originating from the outer edge of the thruster module exit intersects with the reattachment point (schematically illustrated in \cref{fig:ASSchematic}). 
If the pressure ratio is increased beyond its critical value at this flow state, then the trailing Mach wave will intersect with the centerline further downstream of the reattachment point and the wake remains closed. Other than an intuitive reasoning and heuristic evidence, Hagemann does not provide an explanation of the role of the trailing Mach wave in the change of the wake behavior. 
Several experimental and Reynolds-Averaged Navier-Stokes (RANS) investigations show that wake closure can depend on  parameters other than the pressure ratio, such as spike truncation percentage \cite{Nair2019_JSR, Ito2002_JPP} and freestream Mach number \cite{Nair2019_JSR, Takahashi2015_JPP}.
Pyle et al. \cite{Pyle2023_AIAAConf} show that Hagemann's closure criterion may also be met by increasing the jet-height-to-spike-base ratio, $h_i/H$, as defined in  \cref{fig:ASSchematic}. 
A significant gap remains in the understanding of the physical mechanisms and instabilities driving the wake oscillations and related unsteady nozzle loading.
}

\rev{
The  high-resolution, unsteady flow field data that is generated  by high-fidelity computations provides an opportunity to address open questions like these through analysis of linear and non-linear instabilities \cite{Moin1998_ARoF}. 
Spectral solvers, such as discontinuous Galerkin Spectral Element Method (DGSEM) solvers, and high-order finite-difference methods are now frequently used to predict, with high fidelity, the unsteady dynamics and instabilities because of their inherent low dissipation and dispersion errors \cite{Hu1999_JCP, Stanescu2000_JCP, Sherwin1999_PISDGM, Ainsworth2004_JCP, Sengupta2005_AIAAJournal, Sengupta2009_IJMF, Sengupta2009_IJNMF}.
\rev{
With (seemingly) ever-increasing computational resources, direct numerical simulations (DNS) and large eddy simulations (LES) have transitioned from the usage for fundamental flow analysis and simple geometries, such as homogeneous turbulence, channel flow, and backward-facing step flow \cite{Le1997_JFM}, to more complex geometries, such as wing aerodynamics \cite{Klose2020_AIAAJournal, Klose2018_AIAAConf} and scramjet flows \cite{Davidson2005_IJCPD}. 
The simulation of an aerospike nozzle flow, which is characterized by a combination of high-speed, unsteady, viscous flow, shock-vortex interactions, and complex geometry,  challenges even the most robust and accurate numerical methods and Computational Fluid Dynamics software. 
Particularly, the low dissipation of a high-order accurate method leads to spurious numerical oscillations in flow solutions with sharp gradients and strong shocks, whereas a diffusive method may dissipate essential small-scale, flow instabilities. 
Balancing this dissipation in complex flows so that stability is ensured and accuracy is preserved is one of the primary obstacles to conducting simulations of the unsteady aerospike nozzle wake with high-order solvers. 
% Additionally, the computational time required for the simulation of the supersonic and hypersonic aerospike nozzle flows at large Reynolds numbers requires the solver to  be computationally efficient.
Additionally, the significant computational time required for simulating supersonic and hypersonic aerospike nozzle flows at large Reynolds numbers necessitates a computationally efficient solver.
As a result, not many high-fidelity simulations on aerospike nozzle flows have been reported in the literature.  
}
The recent work of Golliard and Mihaescu \cite{Golliard2024_JT} proves the feasibility of computations of aerospike nozzle flows if dissipative finite-volume solvers are employed. However, this approach may overly dissipate the solution and suppress instabilities rather than resolve them.
}

Various shock-capturing schemes have been proposed to improve the stability of spectral solvers while preserving accuracy, such as weighted essentially non-oscillatory (WENO) schemes \cite{Shu2009_SIAMRev, Don2013_JCP}, slope limiters \cite{Cockburn1998_JCP, Flaherty2002_FEAD}, and artificial viscosity (AV) methods \cite{Baldwin1975_ICNMFD, Vonneumann1950_JAP, Abassi2014_CF}. 
\rev{
For a review, we refer the interested reader to the introduction of the article by Chaudhuri et al. \cite{Chaudhuri2017_JCP}. 
In the article, an artificial viscosity method is introduced that scales the numerical dissipation with the viscous and thermal entropy generation according to the entropy conservation law. 
Chaudhuri et al.\cite{Chaudhuri2017_JCP} combine this so-called Entropy Viscosity  (EV) method with an adaptive, element-based, exponential filter that is employed in regions where the artificial numerical viscosity restricts the explicit time step size to satisfy the CFL criterion. 
The viscosity is reduced in viscous regions with large shear gradients using a Ducros switch \cite{Ducros1999_JCP}. 
As robust as this method is, it does not guarantee numerical stability in (hypersonic) flows with strong shocks. 
The Finite Volume Subcell Element (FVSE) scheme introduced by Hennemann et al. \cite{Hennemann2021_JCP} addresses this shortcoming by applying a lower-order finite volume discretization in regions with strong shocks. 
The shock location and strength are identified by the shock-detection method introduced by Persson \cite{Persson2006_AIAAConf} that assesses the modal content of the  polynomial approximation in each element. 
The \FVSE-based shock capturing scheme was shown to be stable for shocks with Mach numbers up to one hundred.
 }
Peck et al. \cite{Peck2024_AIAA} also report that the \FVSE schemes are particularly robust for aerodynamics over a blunt reentry body at Mach 5 compared to the AV  and slope-limiter schemes.
They confirm that AV schemes do not guarantee the positivity and boundedness of pressure and density. 
They also report that the binary nature of limiters leads to temporally inconsistent smoothing and that the residual fails to decay as a result.
In contrast, the \FVSE scheme is reported to be bounded and the solution converges to a steady state.

\rev{
In this paper, we present a discontinuous Galerkin spectral element solver with a hybridization of the Entropy Viscosity (EV) and the Finite Volume Subcell Element (FVSE) shock-capturing scheme to compute the unsteady flow over an  aerospike nozzle.
The \EV method  captures shocks and flow instabilities, while the diffusive \FVSE method is locally called only in regions with strong shocks that can suddenly form in the supersonic jet generated by the aerospike nozzle. A  split-form DGSEM \cite{Klose2020_CF} is employed that numerically preserves kinetic energy.
We will refer to the \DGSEM solver that combines the \EV and \FVSE schemes as the \Hybrid solver. 
The \Hybrid solver is shown to capture strong shocks in the unsteady wake of a perfectly expanded aerospike nozzle flow. 
An underexpanded flow simulation shows that the \Hybrid solver is robust in the hypersonic flow regime. 
A perfectly expanded, three-dimensional, high-speed, turbulent flow simulation at a Reynolds number $Re = 95{,}000$  shows that shocks are captured and the inertial range is resolved in the turbulence energy spectrum.

In the remainder of the paper, we  first summarize the governing equations in \cref{sec:GovEq}. 
The \Hybrid solver is presented in \cref{sec:NumMethods}.
Computational models, including geometry, grid sizing, initial conditions, and boundary conditions are presented in \cref{sec:CaseSetUp}.
A grid independence study is conducted, followed by a discussion of the effect of the numerical method on the flow physics over the aerospike nozzle (\cref{sec:Results}).  
Conclusions are reserved for the last section (\cref{sec:Conclusion}). 

}

\FloatBarrier

\section{Governing Equations}
\label{sec:GovEq}

Aerospike nozzle flows are governed by compressible Navier-Stokes equations. 
In nondimensional vector form, these are
\begin{equation}
    \mathbf{Q}_t + \nabla \cdot {\mathcal{F}} = 0,
    \label{eq:ConsEq_NonDim}
\end{equation}
where $\mathbf{Q}$ is the solution vector,  
\begin{equation}
    \mathbf{Q} = 
    \begin{bmatrix}
        \rho     \quad
        \rho u \quad
        \rho v \quad
        \rho w \quad
        \rho E
    \end{bmatrix}^T,
    \label{eq:NSESolVecDefinition}
\end{equation}
and $\mathcal{F}$ is the flux matrix, 
 \begin{equation*}
     {\mathcal{F}} = 
     \begin{bmatrix}
        \mathbf{F},
        \mathbf{G},
        \mathbf{H}
    \end{bmatrix}.
 \end{equation*}
 $\mathbf{F}, \mathbf{G},$ and $\mathbf{H}$ are the flux vectors in the $x$, $y$, and $z$ directions.
 \rev{
 The flux vectors are the sum of advective and viscous fluxes
 \begin{equation*}
     \mathbf{F} = \mathbf{F}^{a} - \frac{1}{Re_f}\mathbf{F}^{v}, \quad
     \mathbf{G} = \mathbf{G}^{a} - \frac{1}{Re_f}\mathbf{G}^{v}, \quad
     \mathbf{H} = \mathbf{H}^{a} - \frac{1}{Re_f}\mathbf{H}^{v},
 \end{equation*}
 }
where the superscript $a$ denotes the advective fluxes, given by 
\begin{equation}
    \begin{split}
    & \mathbf{F}^{a} =
    \begin{bmatrix*}
        \rho u          \quad\quad
        \rho u u  + p   \quad\quad
        \rho v u        \quad\quad\quad
        \rho w u        \quad\quad\quad\quad\quad
        u (\rho E  + p )
    \end{bmatrix*}^T
    ,\\
    & \mathbf{G}^{a} =
    \begin{bmatrix*}
        \rho  v          \quad\quad
        \rho u v         \quad\quad\quad
        \rho v v   + p   \quad\quad\quad
        \rho w v         \quad\quad\quad\quad
        v (\rho E  + p )
    \end{bmatrix*}^T
    ,\\
    & \mathbf{H}^{a} =
    \begin{bmatrix*}
        \rho  w          \quad\quad
        \rho u w         \quad\quad\quad
        \rho v w         \quad\quad\quad
        \rho w w  + p    \quad\quad\quad
        w (\rho E  + p )
    \end{bmatrix*}^T,
    \end{split}
    \label{eq:NSEFAdvFluxDefinition_NonDim}
\end{equation}
and the superscript $v$ denotes the viscous fluxes, given by
\begin{equation}
    \begin{split}
    & \mathbf{F}^{v} =
    \begin{bmatrix*}[l]
        0         \quad\quad
        \tau_{xx} \quad\quad
        \tau_{yx} \quad\quad
        \tau_{zx} \quad\quad
        u \tau_{xx} + v \tau_{yx} + w \tau_{zx} + \frac{ T_x }{(\gamma - 1)Pr_f M_f^2}
    \end{bmatrix*}^T
    ,\\
    & \mathbf{G}^{v} =
    \begin{bmatrix*}[l]
        0         \quad\quad
        \tau_{xy} \quad\quad
        \tau_{yy} \quad\quad
        \tau_{zy} \quad\quad
        u \tau_{xy} + v \tau_{yy} + w \tau_{zy} + \frac{ T_y }{(\gamma - 1)Pr_f M_f^2}
    \end{bmatrix*}^T
    ,\\
    & \mathbf{H}^{v} =
    \begin{bmatrix*}[l]
        0          \quad\quad
        \tau_{xz}  \quad\quad
        \tau_{yz}  \quad\quad
        \tau_{zz}  \quad\quad
        u \tau_{xz} + v \tau_{yz} + w \tau_{zz} + \frac{ T_z }{(\gamma - 1)Pr_f M_f^2}
    \end{bmatrix*}^T.
    \end{split}
    \label{eq:NSEFViscFluxDefinition_NonDim}
\end{equation}
The total energy density is
\begin{equation}
    \rho E = \frac{p}{\gamma - 1} + \frac{\rho}{2}(u^{2} + v^{2} + w^{2}),
    \label{eq:RhoEDefinition_NonDim}
\end{equation}
and $\tau$ is the shear stress tensor.
The system is closed by the equation of state,
\begin{equation}
    p = \frac{\rho T}{\gamma M_f^2}.
    \label{eq:EqOfState_NonDim}
\end{equation}
The reference Reynolds, Prandtl, and Mach numbers are
\begin{equation*}
    Re_f = \frac{\rho_f^* U_f^* h^*}{\mu^*_f}, \quad Pr_f = \frac{\mu^*(R - c_v)}{\kappa^*}, \quad M_f = \frac{U_f^*}{\sqrt{\gamma R T_f^*}},
\end{equation*}
where the superscript ${}^\star$ denotes a dimensional variable 

\rev{
The temperature-dependent molecular viscosity is computed with the nondimensionalized Sutherland law,
\begin{equation}
    \mu = \frac{\mu^\star}{\mu_f^\star} = (\frac{1 + \mathcal{V}}{T + \mathcal{V}}) T^{3/2}.
\label{eq:Sutherland}
\end{equation}
Here, $\mu_f^\star$ is a reference viscosity, $\mathcal{V} = \frac{\mathcal{V}^\star}{T_f^*}$ is the nondimensionalized Sutherland reference coefficient.
At a reference temperature set to $T_f^*$ = 300 K, it then follows that $\mathcal{V}^\star = 0.368$.
The temperature-dependent thermal conductivity is computed according to:
\begin{equation}
\kappa = \frac{\kappa^\star}{\kappa_f^\star} = \frac{Pr_f}{\mu_f^\star (R - c_v)}  \frac{\mu^\star (R - c_v)}{Pr_f}.
\label{eq:ThermalCond1}
\end{equation}
The nondimensional gas constant, $R$, and specific heat, $c_v$, are assumed constant and \cref{eq:ThermalCond1} reduces to
\begin{equation}
    \kappa = \mu(T) = (\frac{1 + \mathcal{V}}{T + \mathcal{V}}) T^{3/2}.
\label{eq:ThermalCond2}
\end{equation}
}
\FloatBarrier

\section{Numerical Methods}
\label{sec:NumMethods}

\subsection{Discontinuous Galerkin Spectral Element Method (DGSEM)}

The \DGSEM divides the computational domain into $E_{g}$ elements, $\Omega_e$, where $e = 1, 2, \ldots, E_{g}$.
Each element $\Omega_e \in \mathbb{R}^3$ is mapped from the unit cube $\Tilde{\Omega}_e \in [0, 1]^3$ via an isoparametric mapping $\mathbf{x} = \mathbf{X}(\xi, \eta, \zeta)$. 
Under this mapping, 
\cref{eq:ConsEq_NonDim} becomes
\begin{equation}
    \Tilde{\mathbf{Q}}_t + \nabla_{{\xi}} \cdot \Tilde{\mathcal{F}} = 0,
    \label{eq:ConsEq_NonDim_Mapped}
\end{equation}
where the tilde denotes mapped variables and
  $\nabla_\xi \equiv [\partial_\xi, \partial_\eta, \partial_\zeta]^T$.
The solution and flux vectors are approximated by $N^{th}$ order polynomials, e.g.,
\begin{equation*}
    \begin{split}
        \Tilde{\mathbf{Q}}(\xi,\eta,\zeta,t) & \approx \Tilde{\mathbf{Q}}^N(\xi,\eta,\zeta,t) = \sum_{i=0}^N \sum_{j=0}^N \sum_{k=0}^N \mathbf{Q}^N_{ijk}(t) \ell_i (\xi) \ell_j (\eta) \ell_k (\zeta), \\
    \end{split}
    \label{eq:FluxApproxDef}
\end{equation*}
with analogous approximations of $\Tilde{\mathbf{F}}$, $\Tilde{\mathbf{G}}$, and $\Tilde{\mathbf{H}}$.
The Lagrange basis polynomial, 
\begin{equation}
    \ell_i(\xi) = \prod_{\substack{n=0 \\ n \ne i}}^{N} \frac{\xi - \xi_n}{\xi_i - \xi_n},
    \label{eq:LagrangeInterpolationgPoly}
\end{equation}
is defined on Legendre-Gauss-Lobatto quadrature nodes. % and it is thus of the Jacobi family of polynomials.

\rev{
\noindent Substituting the polynomial approximations into \cref{eq:ConsEq_NonDim_Mapped} and taking the inner product with respect to an orthogonal test function $\phi$ leads to the weak form of the governing equations,
\begin{equation}
    \int_\Omega \phi^T(\Tilde{\mathbf{Q}}_t^N + \nabla_\xi \cdot \Tilde{\mathcal{F}}^N) d\Omega = 0.
    \label{eq:InnerProdConsLaw}
\end{equation}
% which minimizes the solution error.
Integrating \cref{eq:InnerProdConsLaw} by parts yields
\begin{equation}
    \int_{\Omega} \phi^T\Tilde{\mathbf{Q}}_t^N d\Omega 
    + \int_{\Gamma} \phi^T (\Tilde{\mathcal{F}}^N)\cdot\hat{\mathbf{n}}d\Gamma 
    - \int_{\Omega} \Tilde{\mathcal{F}}^N \cdot (\nabla \phi)^T d\Omega 
    = 0,
    \label{eq:ConsEq_WeakForm}
\end{equation}
where $\Gamma$ denotes the element surface and $\hat{\mathbf{n}}$ is the outward unit normal.
The connectivity between elements is established through the surface integral in \cref{eq:ConsEq_WeakForm} by replacing the boundary flux at the surface  with a numerical flux $\Tilde{\mathcal{F}}^* (\Tilde{\mathbf{Q}}^{N,-}, \Tilde{\mathbf{Q}}^{N,+})$ that depends on the interface solution of neighboring elements. 
Thus,  
\begin{equation}
    \int_{\Omega} \phi^T\Tilde{\mathbf{Q}}_t^N d\Omega 
    + \int_{\Gamma} \phi^T (\Tilde{\mathcal{F}}^*)\cdot\hat{\mathbf{n}}d\Gamma 
    - \int_{\Omega} \Tilde{\mathcal{F}}^N \cdot (\nabla \phi)^T d\Omega 
    = 0.
    \label{eq:ConsEq_WeakForm_NumFlux}
\end{equation}
The advective numerical fluxes are determined with the Lax–Friedrichs scheme \cite{Rider2002_IJNMF} and the viscous numerical fluxes are evaluated using the Bassi-Rebay (BR1) scheme \cite{Bassi1997_JCP}.
The flux-based connectivity ensures that the approximation is conservative \cite{Gassner2016_JCP}.
}
Integrating \cref{eq:ConsEq_WeakForm_NumFlux} by parts once more leads to the so-called strong form,
\begin{equation}
    \int_{\Omega} \phi^T\Tilde{\mathbf{Q}}^N_t d\Omega 
    + \int_{\Gamma} \phi^T (\Tilde{\mathcal{F}}^* - \Tilde{\mathcal{F}})\cdot\hat{\mathbf{n}}d\Gamma 
    + \int_{\Omega} \phi^T(\nabla \cdot \Tilde{\mathcal{F}}^N) d\Omega 
    = 0.
    \label{eq:ConsEq_StrongForm}
\end{equation}

\noindent 
Following the Galerkin method, we choose the test function to be the same Lagrange 
% Legendre 
polynomial, $\phi = \phi_{ijk} = \ell_i(\xi)\ell_j(\eta)\ell_k(\zeta)$, used for the approximation of the solution vector.
Integrals are evaluated numerically using Legendre-Gauss-Lobatto (LGL) quadrature rules.

% ------------------------

Following the approach of Gassner et al. \cite{Gassner2016_JCP, Gassner2018_JSC} and Klose et al. \cite{Klose2020_CF}, we evaluate the volumetric flux derivatives using split-form, two-point fluxes, so the discretized flux divergence terms become
\begin{equation}
    \begin{split}
        \partial_\xi\Tilde{\mathbf{F}}^N|_{ijk} & \approx 
        (\delta_{ip}[\Tilde{\mathbf{F}}^* - \Tilde{\mathbf{F}}]_{pjk} - \delta_{i0}[\Tilde{\mathbf{F}}^* - \Tilde{\mathbf{F}}]_{0jk}) + 2\sum_{m=0}^N D_{im} \Tilde{\mathbf{F}}^\#_{(i,m)jk}, \\      
         \partial_\eta\Tilde{\mathbf{G}}^N|_{ijk} & \approx 
        (\delta_{jp}[\Tilde{\mathbf{G}}^* - \Tilde{\mathbf{G}}]_{ipk} - \delta_{j0}[\Tilde{\mathbf{G}}^* - \Tilde{\mathbf{G}}]_{i0k}) + 2\sum_{m=0}^N D_{jm} \Tilde{\mathbf{G}}^\#_{i(j,m)k}, \\
         \partial_\zeta\Tilde{\mathbf{H}}^N|_{ijk} & \approx 
        (\delta_{kp}[\Tilde{\mathbf{H}}^* - \Tilde{\mathbf{H}}]_{ijp} - \delta_{k0}[\Tilde{\mathbf{H}}^* - \Tilde{\mathbf{H}}]_{ij0}) + 2\sum_{m=0}^N D_{km} \Tilde{\mathbf{H}}^\#_{ij(k,m)}.
    \end{split}
    \label{eq:SplitForm2PointFluxDeriv}
\end{equation}
\rev{
We compute the advective two-point fluxes using the kinetic-energy-conserving  fluxes proposed by Pirozzoli \cite{Pirozzoli2010_JCP, Pirozzoli2011_ARoF},
\begin{equation}
\Tilde{\mathbf{F}}^{a,\#} =
\begin{bmatrix*}
    \{\{\rho\}\}\{\{u\}\} \\
    \{\{\rho\}\}\{\{u\}\}^2 + \{\{p\}\} \\
     \{\{\rho\}\}\{\{u\}\} \{\{v\}\} \\
      \{\{\rho\}\}\{\{u\}\}\{\{w\}\} \\
       \{\{\rho\}\}\{\{u\}\}\{\{h\}\}
\end{bmatrix*}
, \quad 
\Tilde{\mathbf{G}}^{a,\#} =
\begin{bmatrix*}
    \{\{\rho\}\}\{\{v\}\} \\
     \{\{\rho\}\}\{\{u\}\} \{\{v\}\} \\
     \{\{\rho\}\}\{\{v\}\}^2 + \{\{p\}\} \\
      \{\{\rho\}\}\{\{v\}\}\{\{w\}\} \\
       \{\{\rho\}\}\{\{v\}\}\{\{h\}\}
\end{bmatrix*}
, \quad 
\Tilde{\mathbf{H}}^{a,\#} =
\begin{bmatrix*}
    \{\{\rho\}\}\{\{w\}\} \\
     \{\{\rho\}\}\{\{u\}\} \{\{w\}\} \\
      \{\{\rho\}\}\{\{v\}\}\{\{w\}\} \\
      \{\{\rho\}\}\{\{w\}\}^2 + \{\{p\}\} \\
       \{\{\rho\}\}\{\{w\}\}\{\{h\}\}
\end{bmatrix*}
,
\label{eq:PirozzoliFluxes}
\end{equation}
where the operator $\{\{a\}\}_{im}$ is the two-point average, defined as $\{\{a\}\}_{im} = \frac{1}{2}(a_i + a_m)$, and $h$ is enthalpy.
The Pirozzoli fluxes lead to a kinetic-energy–preserving method and require fewer floating-point operations than other two-point fluxes, such as the Kennedy-Gruber, Ismail–Roe, or Chandrashekar fluxes \cite{Gassner2016_JCP}.
The two-point viscous fluxes are computed as $\mathbf{F}^{v,\#} = \{\{{{\mathbf{F}^v}}\}\}$, $\mathbf{G}^{v,\#} = \{\{{{\mathbf{G}^v}}\}\}$, and $\mathbf{H}^{v,\#} = \{\{{{\mathbf{H}^v}}\}\}$.
Time integration is performed with an explicit fourth-order Runge-Kutta (RK4) scheme. 

The split-form discretization prevents numerical aliasing errors that are  a result of under-integration of the nonlinear volumetric fluxes \cite{Gassner2016_JCP, Gassner2018_JSC}. 
This, in turn, prevents spurious kinetic energy buildup in the solution, and it enhances numerical stability as compared to the standard \DGSEM. 
Klose et al. \cite{Klose2020_CF} assess the stability and accuracy of the split-form \DGSEM for numerous benchmarks of smooth flows and demonstrate
% absolute 
numerical stability and spectral accuracy, even in under- or marginally-resolved flows. 
% These benefits are broadly applicable to nonlinear conservation laws (e.g., Burgers' equation, the compressible Euler equations), not just to Navier-Stokes computations of an aerospike nozzle wake.
The split-form does not, however, prevent numerical oscillations. 
These oscillations are particularly problematic across strong shocks, where undershoots in the solution may result in negative densities and cause numerical instabilities.
For those flow features, we use the \EV and \FVSE shock-capturing scheme to smooth gradients in the solution and ensure boundedness.

}

\subsection{Entropy Viscosity (EV) Shock-Capturing Scheme }
\label{sec:EV}

The entropy viscosity (\EV) scheme introduces an artificial, numerical dissipation term in the governing equations via the viscous fluxes in order to smooth gradients in the solution.
We follow the approaches of Chaudhuri et al. \cite{Chaudhuri2017_JCP}, Chaudhuri \cite{Chaudhuri2019_Entropy}, and Chaudhuri and Jacobs \cite{Chaudhuri2019_SW}, and scale the dissipation term with the viscous and thermal entropy generation. 
We augment the physical viscosity and thermal conductivity with an artificial dissipation component as
\begin{equation}
    \mu_t = {\mu}_{h}Re_{f} + \mu,
\label{eq:combinedAV_mu}
\end{equation}
\begin{equation}
    \kappa_t = {\kappa}_{h}Re_{f} + \kappa,
\label{eq:combinedAV_kap}
\end{equation}
where the subscripts $t$ and $h$ denote total and artificial properties, respectively.
The functions $\mu_{h}$ and $\kappa_{h}$
% are proportional to the viscous and thermal entropy generation, and
are defined as
\begin{equation}
    \mu_{h} = \hat{\mu}_{h} \theta \mathcal{H}(-\nabla \cdot \tilde{\mathbf{u}}),
    \label{eq:AppliedDucrosSensor_mu}
\end{equation}
\begin{equation}
    \kappa_{h}  = {\hat{\kappa}_{h}} \theta,
    \label{eq:AppliedDucrosSensor_kap}
\end{equation}
where $\Tilde{\mathbf{u}}$ is the mapped velocity vector.
We compute $\hat{\mu}_h$ and $\hat{\kappa}_h$ according to 
\begin{equation}
        \hat{\mu}_{h} = C_\mu\frac{\rho(\Delta h)^2}{\| \rho s - \overline{\rho s} \|_\infty}\left[\frac{\Phi}{T}\right],
\label{eq:AVDefinition_mu}
\end{equation}
\begin{equation}
    \hat{\kappa}_{h} = C_\kappa\frac{\rho(\Delta h)^2}{\| \rho s - \overline{\rho s} \|_\infty}\left[\frac{\Gamma}{Pr_{f}(\gamma-1)M_{f}^2T}\right].
\label{eq:AVDefinition_kap}
\end{equation}
\rev{
The functions $\Phi$ and $\Gamma$ denote the viscous and conductive entropy generation and are defined as
\begin{equation}
    \Phi = 2(\mathbf{S} - \frac{1}{3}\mathbf{\delta}\mathbf{S})(\mathbf{S} - \frac{1}{3}\mathbf{\delta}\mathbf{S}),
    \label{eq:ViscEntropyGen}
\end{equation}
\begin{equation}
    \Gamma = \frac{\kappa}{T}\nabla T \cdot \nabla T,
    \label{eq:CondEntropyGen}
\end{equation}
where $\mathbf{S}$ is the strain rate tensor.
The entropy density, $ \rho s$, is given by
\begin{equation}
    \rho s = \frac{\rho}{\gamma (\gamma - 1)M_f^2}\ln\!\left(\frac{p}{\rho^\gamma}\right), 
\end{equation}
where $\|\rho s - \overline{\rho s}\|_\infty$ is the maximum deviation of entropy density from the element average, and $C_\mu$ and $C_\kappa$ have a unity order of magnitude.
The minimum nodal spacing is $\Delta h$.
}
The Ducros sensor \cite{Ducros1999_JCP, Pirozzoli2011_ARoF, Chaudhuri2017_JCP}, 
\begin{equation}
    \theta = \frac{(\nabla\cdot \tilde{\mathbf{u}})^2}{(\nabla\cdot \tilde{\mathbf{u}})^2 + (\nabla \times \tilde{\mathbf{u}})^2 + \epsilon},
    \label{eq:DucrosSensor}
\end{equation}
reduces the artificial dissipation in  smooth (turbulent) viscous shear dominated flows ($\theta \to 0$) while maintaining \EV dissipation near dilatation dominated flow regions, i.e., shocks ($\theta \to 1$). 
Here, $\epsilon = 10^{-12}$  prevents division by zero.
The Heaviside step function, $\mathcal{H}$,  in \cref{eq:AppliedDucrosSensor_mu} sets the artificial viscosity to zero in regions with isentropic, expanding flows. 
%near shocks and $\theta \to 0$ in shear-dominated regions.
% This prevents over-dissipation in regions of high entropy generation and physical viscosity (e.g., boundary and shear layers).
The \EV formulation was shown to provide high-order resolution  in  grid-resolved, smooth flow regions \cite{Chaudhuri2017_JCP, Abassi2014_CF}.

A drawback of artificial viscosity methods is its constraint on the viscous, stable time step if an  explicit time integration scheme is used.
As the artificial viscosity increases, the explicit, stable, viscous time step,
\begin{equation}
    \Delta t_{v} = CFL_v Re_f\frac{(\Delta h)^2}{\nu + \nu_{h}},
    \label{eq:ViscTimeStep}
\end{equation}
becomes prohibitively small.
To mitigate this, Chaudhuri et al. \cite{Chaudhuri2017_JCP} restrict $\mu_h$ to
\begin{equation}
    \mu_h \le \mu_{max} = C_m \rho \Delta h (\| \mathbf{u} \| + \sqrt{T}).
    \label{eq:MuAVMax}
\end{equation}
The parameter $C_m$ has a unity order of magnitude.  
No analytical bounds on $(C_\mu, C_\kappa, C_m)$ have been derived that ensure numerical stability. These parameters may vary slightly on a case by case basis.
Chadhuri et al. \cite{Chaudhuri2017_JCP} used an element-based, exponential filter in regions where the entropy viscosity is restricted according to this stable, explicit time step. 
The element-based filter was also proposed for an explicitly filtered Large-Eddy Simulation approach by Sengupta et al. \cite{Sengupta2009_IJNMF}. 
We do not supplement the \EV with a filter in the \Hybrid scheme.
Rather, we adaptively blend a dissipative \FVSE solution with the \DGSEM solution using an \EV-based shock indicator.
The \FVSE solution is described in \cref{sec:FVSE}, and the \EV-based shock indicator and resultant blending coefficient is discussed in \cref{sec:HybridScheme}.

\subsection{Finite Volume Subcell Element (FVSE) Shock-Capturing Scheme }
\label{sec:FVSE}

\begin{figure}[ht]
    \centering
    \includegraphics[width=0.45\linewidth]{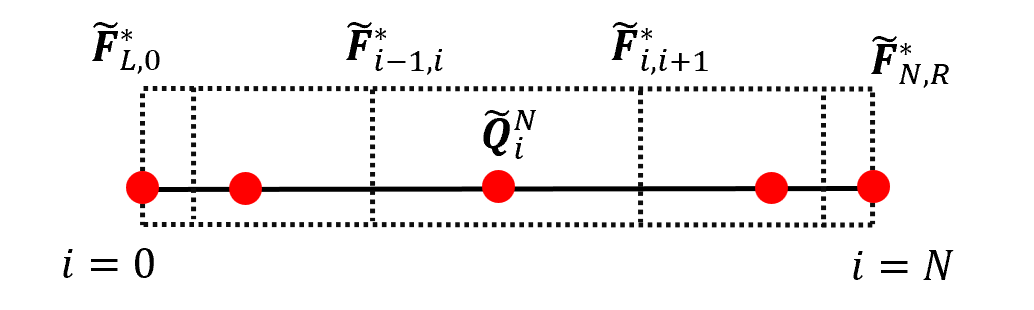}
    \caption{ \rev{A 1D spectral element (solid line) and corresponding finite volume subcell elements (dashed lines). $\Tilde{\mathbf{Q}}^N$ is the subcell average and $\Tilde{\mathbf{F}}^*$ is the interface flux.}
    }
    \label{fig:FVSEIllustration}
\end{figure}

The \FVSE scheme, as discussed by Hennemann et al. \cite{Hennemann2021_JCP}, is a dissipative, low-order, finite volume discretization that is applied within a \DGSEM element. 
\rev{
The \DGSEM element is subdivided into subcells that are centered at the quadrature nodes, which define the location of the subcell average.
Accordingly, each \DGSEM element has  $(N+1)^{d}$ subcells as illustrated for $d$=1 in \cref{fig:FVSEIllustration}.
Here, $d$ denotes the spatial dimension.
}

The subcell averages are  initialized with the \DGSEM solution at the quadrature nodes at each RK stage.
The inviscid fluxes at the subcell interfaces, which lie between the quadrature nodes, are determined with the local Lax-Friedrichs (\LLF) scheme \cite{Rusanov1962_USSR, Toro, Leveque}
\rev{
\begin{equation}
    \Tilde{\mathcal{F}}^*_{i+1/2} = \frac{1}{2}( \Tilde{\mathcal{F}}_{i} +  \Tilde{\mathcal{F}}_{i+1}) - \frac{\max{(\lambda_{i},\lambda_{i+1})}}{2}(\Tilde{\mathbf{Q}}_{i+1} - \Tilde{\mathbf{Q}}_{i}),
\label{eq:LxFFlux}
\end{equation}
with an analogous stencil between nodes $i-1$ and $i$.
Here, $\max{(\lambda_{i},\lambda_{i+1})}$ denote the maximum wavespeed at nodes $i$ and $i+1$.
The viscous fluxes are computed with the BR1 scheme.
Under this discretization, numerical dissipation is introduced via the \LLF jump term [\cref{eq:LxFFlux}] in the inviscid flux computation.
The resulting \FVSE solution within subcell $ijk$ of a \DGSEM element is computed according to
\begin{equation}
    (\Tilde{\mathbf{Q}}^N_t)_{ijk}^{FVSE} + (\nabla_\xi \cdot \Tilde{\mathcal{F}}^N)_{ijk}^{FVSE} = 0.
\label{eq:FVSESol}
\end{equation}

Hennemann et al. \cite{Hennemann2021_JCP} show that the \FVSE and \DGSEM solutions may be blended to increase a solver's numerical dissipation \cite{Hennemann2021_JCP}.
If this blending is coupled with a shock indicator, it can be localized to shocked elements and understood as an \FVSE-based shock-capturing scheme.
We propose a unique \EV-based blending in \cref{sec:HybridScheme}.

}
%%%%%%%%%%%%%

\subsection{Hybrid Shock-Capturing Scheme}
\label{sec:HybridScheme}

\begin{figure}[ht]
    \centering
    \includegraphics[width=0.385\linewidth]{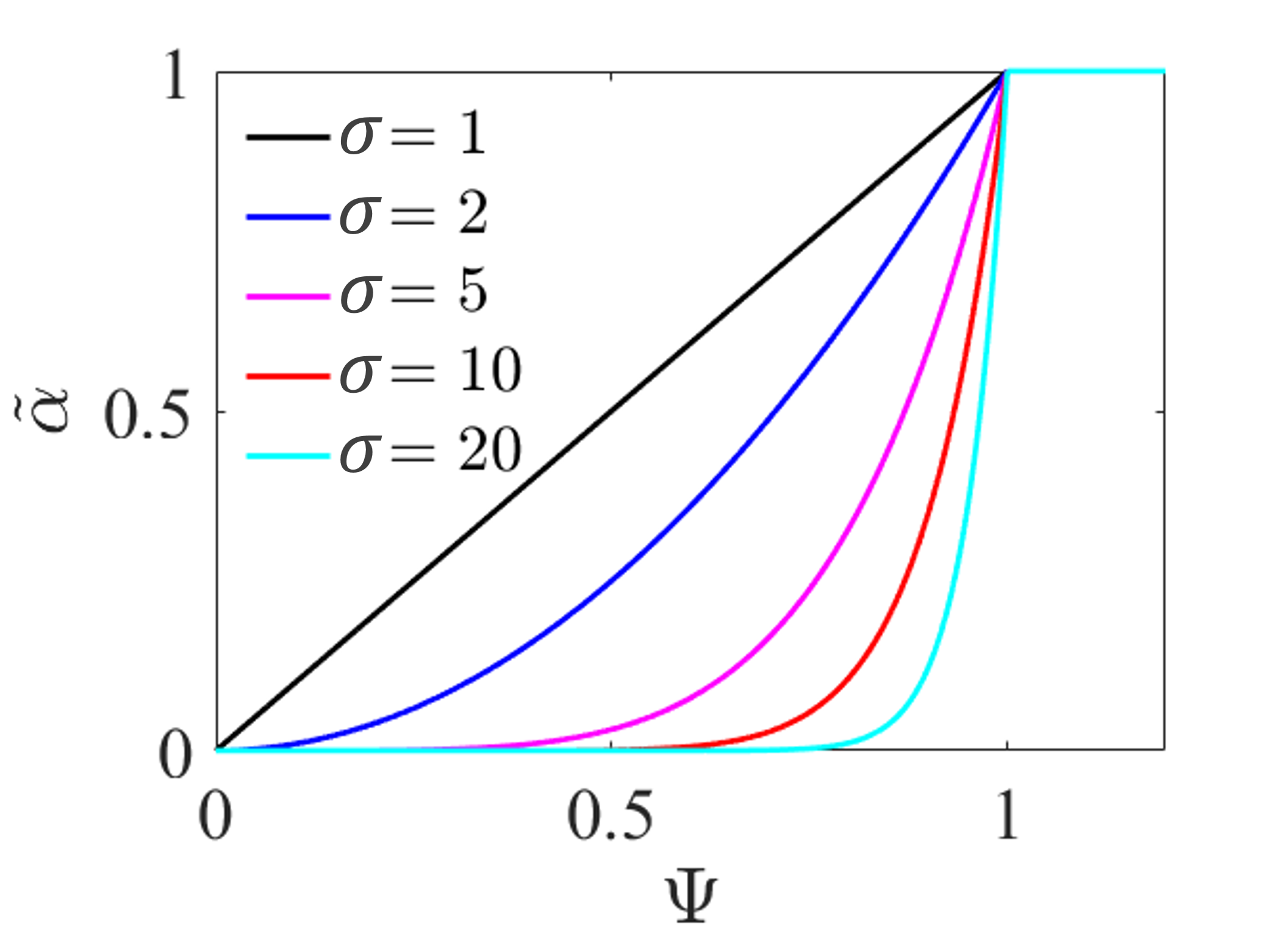}
    \caption{The \FVSE blending coefficient $\Tilde{\alpha}$ as a function of the shock sensor $\Psi$. 
    }
    \label{fig:AlphaTildePlot}
\end{figure}

We hybridize the high-order, \EV-stabilized \DGSEM scheme and the low-order, dissipative \FVSE scheme according to the linear blending function
\begin{equation}
    (\Tilde{\mathbf{Q}}^N_t)_{e} + \alpha (\nabla_\xi \cdot \Tilde{\mathcal{F}}^N)^{FVSE}_e + (1-\alpha)(\nabla_\xi \cdot \Tilde{\mathcal{F}}^N)^{DGSEM}_e = 0.
\label{eq:ConsEq_Residual}
\end{equation}
We blend the numerical solution vector on a per-element basis, as indicated by the subscript "$e$", using the element-based blending coefficient, $\alpha$.
Hennemann et al. show that an element-wise blending, rather than node-wise, between \FVSE and \DGSEM solutions preserves local conservation.
The hybridization adaptively adds numerical dissipation through the inviscid fluxes, via the \LLF fluxes used in the \FVSE scheme, and the viscous fluxes, via the \EV dissipation terms [\cref{eq:combinedAV_mu,eq:combinedAV_kap}].
% If the \FVSE-\DGSEM blending is localized to shock-containing elements (see, for example, \cite{Hennemann2021_JCP, Peck2024_AIAA}), 
The numerical dissipation introduced by the \FVSE solution functions as shock-capturing scheme that is supplementary to the \EV shock-capturing scheme. We couple \FVSE and \EV schemes by determining the \FVSE blending coefficient $\alpha$ using the \EV-based shock indicator,
\begin{equation}
    \Psi_{ijk} = \max\left(\frac{\mu_h}{\mu_{max}}, \frac{\kappa_h}{\mu_{max}}\right)_{ijk}.
\label{eq:PsiEq}
\end{equation}
$\Psi$ is computed per-node within the spectral element based on the maximum of the normalized momentum and thermal conductivities at each node.
The \FVSE blending coefficient is related to this shock indicator according to
\begin{equation}
    \Tilde{\alpha}_{ijk} =  \Psi^\sigma_{ijk} + \alpha_{min},
\label{eq:BlendingFunction_AlphaTilde}
\end{equation}
where $\Tilde{\alpha}_{ijk} \in [0, 1]$, $\sigma$ denotes an integer power of one or higher, and $\alpha_{min}$ is a minimum global blending coefficient.
Increasing $\sigma$ steepens the $\Tilde{\alpha}(\Psi)$ trend, widening the $\Psi$ band in which $\Tilde{\alpha}(\Psi) \approx \alpha_{min}$ (\cref{fig:AlphaTildePlot}).
% This steepening reduces the \FVSE weighting of the hybrid scheme [\cref{eq:ConsEq_Residual}] for a given $\Psi$.

Following Hennemann et al. \cite{Hennemann2021_JCP}, we restrict the \FVSE weighting in \cref{eq:ConsEq_Residual} by enforcing $\alpha_{ijk} \in [\alpha_{min}, \alpha_{max}]$ through the piecewise function
\begin{equation}
    \begin{split}
    \alpha_{ijk} = 
    \begin{cases}
        \alpha_{min}                     \quad &, \quad \Tilde{\alpha}_{ijk} < \alpha_{min} \\
        \Tilde{\alpha}_{ijk}                    \quad &, \quad \alpha_{min} \le \Tilde{\alpha}_{ijk} \le \alpha_{max} \\
        \alpha_{max}                     \quad &, \quad \Tilde{\alpha}_{ijk} > \alpha_{max} \\
    \end{cases}
    ,
    \end{split}
\label{eq:BlendingFunction_Alpha}
\end{equation}
% For example, if $\alpha_{max} = 0.1$, the hybrid solution will be comprised of, at most, $10\%$ \FVSE solution and at minimum $90\%$ \DGSEM solution.
Finally, we determine the element-wide blending coefficient to be the maximum $\alpha$ within the element:
\begin{equation}
    \alpha = \alpha_m = \max(\alpha_{ijk})_e.
\label{eq:AlphaM}
\end{equation}

\rev{
Because the \FVSE blending coefficient is determined from the \EV-based shock indicator, $\alpha$ depends indirectly on the \EV coefficients $(C_\mu, C_\kappa, C_m)$ through \cref{eq:AVDefinition_mu,eq:AVDefinition_kap}.
The coefficients $C_\mu$ and $C_\kappa$ directly scale $\mu_h$ and$\kappa_h$, thereby increasing both \EV dissipation [\cref{eq:AVDefinition_mu,eq:AVDefinition_kap}] and the \FVSE weighting [\cref{eq:PsiEq,eq:BlendingFunction_AlphaTilde}].
In contrast, $C_m$ has no direct effect on $\mu_h$ or $\kappa_h$ but may affect $\Psi$, and thus $\alpha$, through the $\mu_{max}$ normalization in \cref{eq:BlendingFunction_AlphaTilde}.
We examine the effects of the $C_m$ coefficient and $\alpha$ on the \Hybrid solution in \cref{sec:Results}.

}

\FloatBarrier

\section{Computational Models}
\label{sec:CaseSetUp}

\rev{
We present four computational models for testing and verification of the \Hybrid scheme:
a subsonic, square cylinder wake,
perfectly expanded (2D) aerospike nozzle flow,
underexpanded (2D) aerospike nozzle flow,
and perfectly expanded (3D) aerospike nozzle flow.
We discuss the computational domain and grid sizing, then present initial and boundary conditions for each model.
}

\rev{
\subsection{Square Cylinder Flow: Convergence and Computational Efficiency  }

We follow the computational setup of the subsonic square cylinder wake as described by Klose et al. \cite{Klose2020_CF}.
To summarize, the computational domain spans  $x\in [-1, 15]$, $y\in [-3, 3]$ and
the square cylinder is located at  $x\in [0, 1]$, $y\in [-0.5, 0.5]$.
The cylinder surfaces are no-slip, adiabatic walls.
The freestream Mach and Reynolds numbers are set to $M=0.1$ and $Re_f=100$, respectively. 
The solution is integrated over $200$ convective time units. 
Time averages are taken over the time interval $t\in[100,200]$. 
We conduct a grid-independence study using orthogonal grids with uniform spacings of $\Delta x = \Delta y = \Delta s$, and $\frac{1}{\Delta s} = 4, 8, 16, 32$. 
The polynomial order for the approximation of the solution vector is $N$=3. 
}

\subsection{Aerospike Nozzle Flow: Two-Dimensional Perfectly Expanded Flow}
\label{sec:ThrusterMach2NonExp}

We consider two-dimensional, perfectly expanded flow over a linear, truncated aerospike nozzle with straight (non-contoured) ramps in quiescent air at sea level.
The geometry is representative of a space plane propulsion system, such as the one used in POLARIS's MIRA and MIRA II vehicles \cite{PolarisArticle2023, PolarisArticle2023_2}, and has been the subject of numerous experimental \cite{Mueller1971_UniversityOfND, Mueller1972_EF, Mueller1972_UniversityOfND, Mueller1973_ASME, Verma2009_PropConf, Verma2011_JPP, Takahashi2015_JPP, Sieder2024_A} and computational studies \cite{Hagemann1998_JPP, Chutkey2012_JSR, Soman2021_JSR, Chai2021_AIAAConf, Pyle2023_AIAAConf, Jency2025_Aerospace}. 

\begin{figure}[ht]
    \centering
    \includegraphics[width=0.45\textwidth]{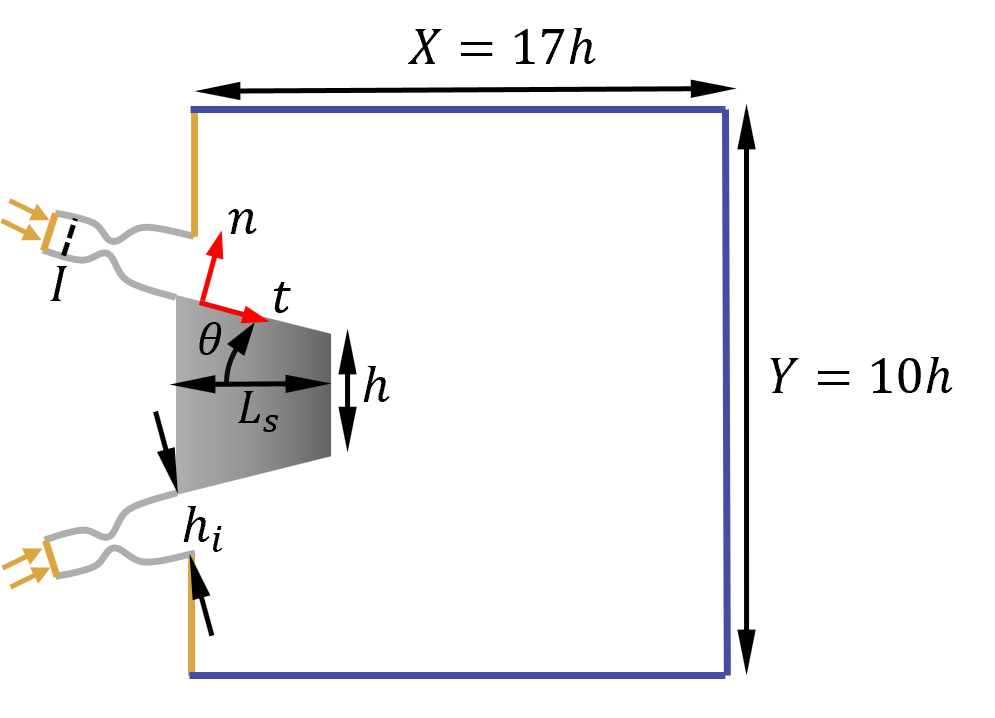}
    \caption{ The computational domain of the two-dimensional aerospike test cases.
    The domain size is not to scale. 
    }
\label{fig:ASIC2D}
\end{figure}

The ramp angle is $\theta = 15^\circ$, and the non-dimensional spike length is $L_s = 0.8571$ (\cref{fig:ASSchematic}). 
The geometry of the supersonic section of the thruster module is designed according to an isentropic minimum-length supersonic nozzle \cite{Anderson}. 
The corresponding wall contour is determined using the method of characteristics  \cite{Pyle2023_AIAAConf}. 
The subsonic section of the thruster module is specified according to a cosine function, as done in \cite{Cui2015_IWMECS},
\begin{equation}
    g(x) = \frac{y_{I} - y_{th}}{2} (1 + \cos(\pi\frac{x - x_{I}}{x_{th} - x_{I}})) + y_{th}.
\label{eq:subsonicNozCont}
\end{equation}
Variables $(x_{I}, y_{I})$ are the coordinates of the subsonic inlet (\cref{fig:ASIC2D}), and $(x_{th}, y_{th})$ are the coordinates of the location of the throat.
A straight section of length $0.25$ is prepended to the subsonic contour.
Without the prepended section, spurious wave reflections are observed at the inlet before the thruster flow achieves a steady state.

\begin{figure}[ht]
    \centering
    \includegraphics[width=0.33\textwidth]{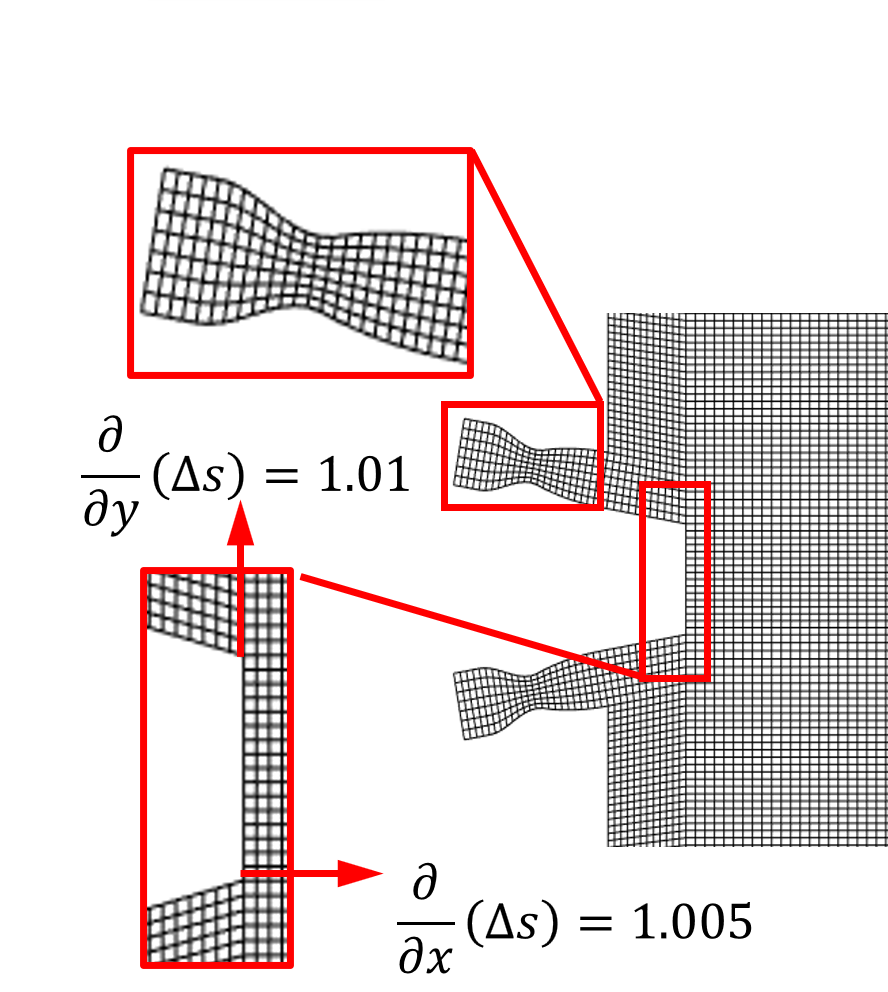}     
    \caption{The two-dimensional, computational grid.
    The domain and element sizes are not to scale.
    }
    \label{fig:GridExample}
\end{figure}

The computational domain spans $x \in [-1.26, 17]$, $y \in [-5, 5]$.
A uniform grid spacing is set with $\Delta x = \Delta y = \Delta s = 1/64$ along the spike ramps and the base of the aerospike nozzle.
The grid spacing increases linearly in the positive $x$ direction and away from the centerline in the $y$ direction, as visualized in \cref{fig:GridExample}. 
We impose a maximum aspect ratio $\Delta x / \Delta y \le 1.1$ for the elements in the region $x \in [-1.26, 15], y\in [-3, 3]$.
Sharp gradients (i.e., shocks) are present in both the $x$ and $y$ directions in this wake region and one can therefore not stretch the grid in a direction with smaller gradients.
% Larger aspect ratios led to numerical instabilities because shocks and sharp gradients are present in both the $x$ and $y$ directions in this wake region, and one can therefore not stretch the grid in a direction with smaller gradients.
The restriction on the aspect ratio is relaxed to $\Delta x / \Delta y \le 3$ further from the wake, in the region $x \in [15, 17]$, $|y| \in [3, 5]$.
The order of approximation is $N = 3$.

The wall-bounded jet flow is initialized in the direction normal to the ramp, i.e., in the $t-n$ coordinate frame (\cref{fig:ASIC2D}), such that
\begin{equation}
    \Tilde{\mathbf{Q}}^N = 
    \begin{cases}
        \Tilde{\mathbf{Q}}^N_{I} & t \le t_{I} \\
        \Tilde{\mathbf{Q}}^N_a & t > t_{I} \\
    \end{cases}
    ,
    \label{eq:InletCond}
\end{equation}
where subscripts $I$ and $a$ denote the inflow and ambient states.
We compute $t_{I}$ as 
\begin{equation}
    \begin{bmatrix}
        n_{I} \\
        t_{I}
    \end{bmatrix}
    = B(\theta)
    \begin{bmatrix}
        x_{I} \\
        y_{I}
    \end{bmatrix}
    ,
\label{eq:TCComp}
\end{equation}
where $B(\theta)$ is the rotation matrix and $\theta$ is the ramp angle.
Using \cref{eq:TCComp}, we compute $t_n = -1.4$.
All inlet and outlet conditions, highlighted by yellow and blue boundaries in \cref{fig:ASIC2D}, are specified according to \cref{eq:InletCond}.
The uniform  inflow condition is specified according to the inlet Mach number, $M_{I} = 0.3$.
We find this inlet condition more robust than an inflow condition that specifies a stagnation state.

The flow expands from $M_{I} = 0.3$ to an exit Mach number of $M_i = 2$ through the thruster module.
The pressure at the thruster exit plane is equal to the ambient pressure, $p_i = p_a$, so the flow over the aerospike ramp is perfectly expanded. 
Following Chai et al. \cite{Chai2021_AIAAConf} and Pyle et al. \cite{Pyle2023_AIAAConf}, we impose $\rho_i = \rho_a$ and $T_i = T_a$, and the stagnation state follows from the isentropic relations. 
We summarize the specification of the velocity and thermodynamics state variables at stagnation ("o"), the thruster inlet ("I") and exit ("i"), and the ambient ("a") in \cref{tab:ThrusterMach2NonExpVarValues}.
Reference variables ("f") are included for readers interested in the dimensional flow specifications.
The Reynolds number based on the height of the spike and the velocity at the thruster exit is $Re_f=95,000$.
For this nozzle flow, we expect instabilities to develop in the shear layers and the wake \cite{Ragab1989_PF, Bashkin2000_FD, Harris1998_AIAA} that are not captured in comparable RANS computations \cite{Pyle2023_AIAAConf, Chai2021_AIAAConf, Soman2021_JSR, Nair2017_JTS, Jency2025_Aerospace}.

% This ensures future investigations may study instabilities along the outer shear layer (\cref{fig:ASSchematic}) independent of screech tones or Mach waves found in overexpanded and underexpanded jets.
% Screech tones excite shear instabilities in jets \cite{Tam1989_JFM, Tam1995_ARoF} and enhance individual modes, which can obscure the dominant frequencies of the non-excited wake \cite{Golliard2024_JT}.
% Expansion - compression cells formed by rebounding Mach waves may coalesce into oblique shocks and cause flow separation \cite{Pyle2023_AIAAConf}.

\begin{table}[ht]
\centering
\caption{ Flow velocity magnitude ($U$), thermodynamic state variables ($\rho, T, p$), and Mach number at stagnation ("o"), the thruster inlet ("I") and exit ("i"), and the ambient ("a") for the perfectly expanded Mach 2 case. 
Reference values are included in row "f".
}
\begin{tabular}{cccccc}
    \hline
    \hline
    $  $  &  $\rho$            &  $U$           &  $T$           &  $p$               & $M$  \\
    \hline 
    Stagnation ($o$)  &  $4.35$            &  $0$           &  $0.45$        &  $1.40$            & $0$   \\
    %\hline
    Thruster inlet ($I$)  &  $4.18$            &  $0.19$        &  $0.44$        &  $1.31$            & $0.3$   \\
    %\hline
    Thruster outlet ($i$)  &  $1$               &  $1$           &  $0.25$        &  $0.18$            & $2$   \\
    %\hline
    Ambient ($a$)  &  $1$               &  $0$           &  $0.25$        &  $0.18$            & $0$   \\    
    %\hline
    Reference ($f$)  &  $1.24$ $[kg/m^3]$ &  $676$ $[m/s]$ &  $1137$  $[K]$ &  $576,288$ $[Pa]$  & $1$   \\
    \hline
    \hline
\end{tabular}
\label{tab:ThrusterMach2NonExpVarValues}
\end{table}

\rev{
All spike walls are treated as slip walls.
We deliberately do not resolve the turbulent boundary layer over the thruster and aerospike walls.
We have found, using RANS simulations and a method of characteristics (MoC) solver\cite{Pyle2023_AIAAConf}, that the slip wall boundary is sufficient to capture the wake flow physics.
Moreover, wall-resolved LES / DNS necessitates a wall-normal grid spacing of $\Delta y^+ < 1$ and sufficient resolution in the log-layer to resolve all turbulent scales.
Kim et al. \cite{Kim1987_JFM} and Moin and Mahesh \cite{Moin1998_ARoF} report grid spacings of $\Delta x^+ \approx 12$ and $\Delta z^+ \approx 7$ are required to fully resolve the turbulent boundary layer in the $x$ and $z$ directions.
Here, $\Delta x^+$, $\Delta y^+$, and $\Delta z^+$ are grid spacings in the $x$, $y$, and $z$ directions, normalized by the viscous length scale $\frac{\nu}{u_\tau}$.
For a conservative estimate of the resolution requirements, we assume a uniform wall-normal spacing of $\Delta y^+ = 0.8$.
Using the turbulent boundary correlations discussed by Schlichting \cite{Schlichting} to estimate the viscous Reynolds number, we compute $Re_\tau = 428$ at the spike tip for the perfectly expanded Mach $2$ aerospike case.
We follow Choi et al. \cite{Choi2012_PoF} and scale the $\Delta x^+$, $\Delta y^+$, $\Delta z^+$ by $Re_\tau$. 
These resolution requirements yield an additional $1140$, $430$, and $750$ nodes in the $x$, $y$, and $z$ directions to resolve the boundary layer along one thruster - aerospike wall.  
Thus, resolving the boundary layer would require $\mathcal{O}(1.8\times10^6)$ and $\mathcal{O}(2.9\times10^9)$ elements in two and three dimensions, respectively, rendering three-dimensional studies impractically expensive.
Of course, we may reduce the near-wall resolution requirements by decreasing $Re_f$.
However, DNS of a supersonic plane wake by Harris and Fasel \cite{Harris1998_AIAA} shows that doing so can suppress  essential instabilities along the separated shear layer that perturbs the reattachment shocks.
}

Because the flow through expansion fans is isentropic, the entropy viscosity is close to zero ($\hat{\mu}_h \rightarrow 0$), thus neither \EV nor \FVSE methods can stabilize the solution here. However, the  singularity of a corner expansion is well-known to lead to numerical instabilities (see, for example, \cite{Woodward1984_JCP}).
Therefore, we increase the blending coefficient $\alpha$ in the expanding flow at the spike corners as follows:
% \begin{equation}
%     \alpha = 
%     \begin{cases}
%         1,      & \sqrt{(x - x_{cr})^2 + (y - y_{cr})^2} < R_{cr} \\
%         \Psi^\sigma + \alpha_{min}, & \sqrt{(x - x_{cr})^2 + (y - y_{cr})^2} > R_{cr}
%     \end{cases}
%     ,
% \label{eq:CornerAlpha}
% \end{equation}
\begin{equation}
    \alpha = 
    \begin{cases}
        1,      & \sqrt{(x - x_{cr})^2 + (y - y_{cr})^2} < R_{cr} \\
        \alpha_m, & \sqrt{(x - x_{cr})^2 + (y - y_{cr})^2} > R_{cr}
    \end{cases}
    ,
\label{eq:CornerAlpha}
\end{equation}
where $(x_{cr}, y_{cr})$ are the coordinates of the spike corners and $R_{cr}$ is a prescribed radius.
Here, $\alpha_m$ is the element-maximum $\alpha$ defined in \cref{eq:AlphaM}.
The explicitly assigned $\alpha$= 1 region ensures boundedness of the solution. 
% across the Prandtl-Meyer fans at the spike corners.
For $t < 3$, $R_{cr} = 0.2$ and for $t \ge 3$, $R_{cr} = 0.0125$.
We also assign $\alpha = 1$ in the thruster modules to ensure numerical stability as a shock front develops.

We impose a buffer or damping layer in the region $x \in [15, 17]$, $|y| \in [3, 5]$ to prevent spurious boundary reflections \cite{Jacobs2003_NHT}.
For $x \in [15, 16]$, $y \in [3, 4]$, $\alpha$ is increased linearly according to 
\begin{equation}
    \alpha = \max \biggl(\frac{x - x_{st}}{x_{sp} - x_{st}},  \frac{y - y_{st}}{y_{sp} - y_{st}}\biggr).
\label{eq:FVSEDampingLayer}
\end{equation}
In this case, $x_{st} = 15$, $x_{sp} = 16$, $y_{st} = 3$, and $y_{sp}  = 4$.
$\alpha = 1$ in the range $x \in [16, 17]$, $y \in [4, 5]$.

\begin{figure}[ht]
    \centering
    \begin{subfigure}[b]{0.28\textwidth}
        \includegraphics[width=\textwidth]{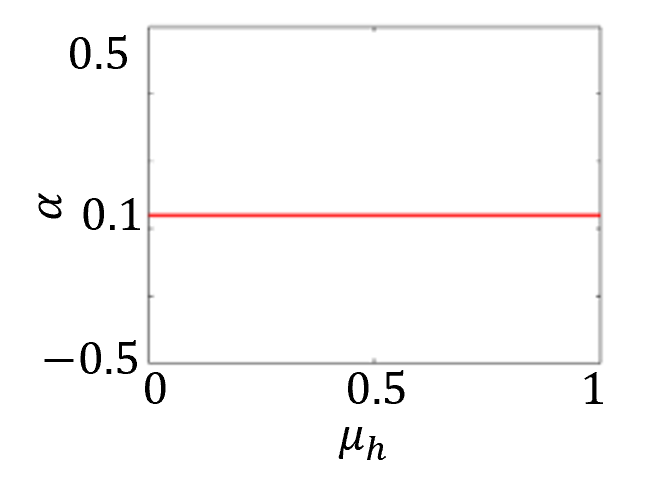}
        \caption{ }
        \label{fig:AlphaVsMu_1}
    \end{subfigure} 
    \begin{subfigure}[b]{0.28\textwidth}
        \includegraphics[width=\textwidth]{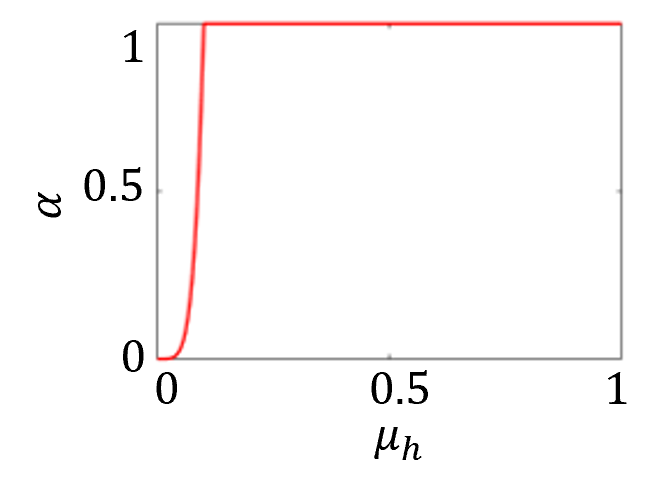}
        \caption{ }
        \label{fig:AlphaVsMu_2}
    \end{subfigure} 
    \begin{subfigure}[b]{0.28\textwidth}
        \includegraphics[width=\textwidth]{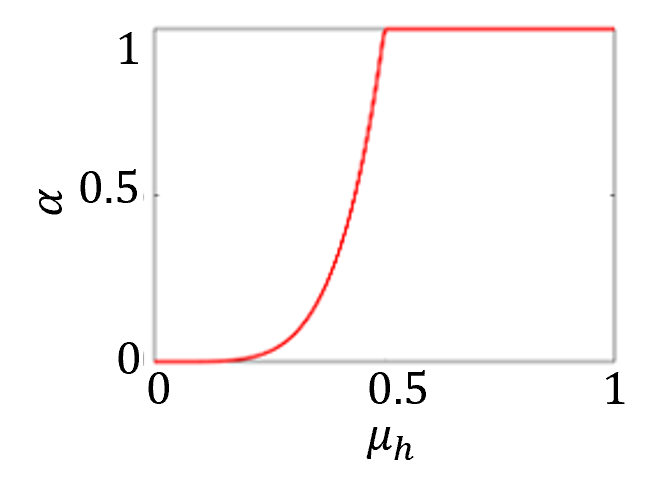}
        \caption{ }
        \label{fig:AlphaVsMu_3}
    \end{subfigure} 
    \caption{The blending coefficient as a function of $\mu_h$ in the (a) A01, (b) CM01, and (c) CM05 models.
    }
    \label{fig:AlphaVsMuExamples}
\end{figure}

We compare three blending models to investigate the effects of the \FVSE blending coefficient $\alpha$ and the \EV coefficient $C_m$ on the flow field.
The first blends the \FVSE scheme using a constant $\alpha = 0.1$ in all elements, independent of $\Psi$.
We refer to this as the "A01" model.
In the second, $\alpha$ is computed using the \EV-based shock indicator $\Psi$.
The maximum artificial viscosity, $\mu_{max}$, used to normalize $\mu_h$ in \cref{eq:PsiEq} is computed according to \cref{eq:MuAVMax} with $C_m = 0.1$.
We refer to this as the "CM01" model.
In the third model, we increase the \EV coefficient to $C_m=0.5$ in order to increase $\mu_{max}$.
As a result, the shock indicator $\Psi$, and thus the blending coefficient $\alpha$, is reduced [\cref{eq:PsiEq,eq:BlendingFunction_AlphaTilde}] and less numerical dissipation is introduced through the \FVSE blending as compared to the CM01 case.
We refer to this as the "CM05" model.
In all cases, $\sigma = 5$ (Eq. \ref{eq:BlendingFunction_AlphaTilde}), $\alpha_{max} =1$, and $\alpha_{min} = 0$ [\cref{eq:BlendingFunction_Alpha}].
We plot $\alpha(\mu_h)$ in \cref{fig:AlphaVsMuExamples} to illustrate the hybridized \EV-\FVSE shock-capturing scheme in each case.

\rev{
To develop a general understanding of how the \Hybrid scheme behaves relative to validated shock-capturing schemes, we compare the A01, CM01, and CM05 models to the entropy viscosity shock-capturing scheme presented by Chaudhuri et al. \cite{Chaudhuri2017_JCP}.
We refer to this as the "EVF" scheme.
As discussed in the introduction, we apply an exponential filter in elements where $\mu_h \ge \mu_{max}$ in this scheme.
Unlike the \FVSE scheme, the EVF scheme does not guarantee pressure and density positivity of the solution in regions with strong shocks, i.e., where $\mu_h > \mu_{max}$.
Chai et al. \cite{Chai2021_AIAAConf} report that the EVF scheme stabilizes aerospike nozzle simulations over short times, however, the sudden formation of shocks in the wake and resulting multi-shock interactions led to numerical instabilities before statistical convergence could be achieved.
% Indeed, we will show that overshoots and undershoots in the solution across a shock are present if the EVF scheme is used, but not the \Hybrid scheme.
For further details on the exponential filter, we refer to \cite{Chaudhuri2017_JCP} and references therein.
}

\FloatBarrier

\subsection{Aerospike Nozzle Flow: Two-Dimensional Underexpanded Flow}

To assess the performance of the \Hybrid scheme for hypersonic conditions, we simulate a two-dimensional, underexpanded aerospike nozzle flow.
The ambient pressure is reduced to expand the flow over the aerospike nozzle from $M = 2$ to $M = 5$, although we will demonstrate that Mach numbers up to $M = 6$, i.e., hypersonic conditions,  are attained.
The resulting shocks are stronger and the solution gradients steeper than in the perfectly expanded (Mach $2$) case.
Hence, this case provides a rigorous test of the scheme’s ability to maintain monotonicity and boundedness in solutions with sharp gradients.
The flow conditions are summarized in \cref{tab:ThrusterMach2ExpVarValues}.
Aside from the ambient conditions, all other flow variables are the same as for the perfectly expanded nozzle flow (\cref{tab:ThrusterMach2NonExpVarValues}). 
We use the CM05 model.

\begin{table}[ht]
\centering
\caption{ Flow velocity magnitude ($U$), thermodynamic state variables ($\rho, T, p$), and Mach number at stagnation ("o"), the thruster inlet ("I") and exit ("i"), and the ambient ("a") for the underexpanded Mach 2 case. 
Reference values are included in row "f".
}
\begin{tabular}{cccccc}
    \hline
    \hline
    $ $                   &  $\rho$            &  $U$           &  $T$           &  $p$               & $M$  \\
    \hline 
    Stagnation ($o$)      &  $4.35$            &  $0$           &  $0.45$        &  $1.40$            & $0$   \\
    %\hline
    Thruster inlet ($I$)  &  $4.18$            &  $0.19$        &  $0.44$        &  $1.31$            & $0.3$   \\
    %\hline
    Thruster outlet ($i$) &  $1$               &  $1$           &  $0.25$        &  $0.18$            & $2$   \\
    %\hline
    Ambient ($a$)         &  $0.05$            &  $0$           &  $0.08$        &  $0.0026$          & $0$   \\    
    %\hline
    Reference ($f$)       &  $1.24$ $[kg/m^3]$ &  $676$ $[m/s]$ &  $1137$  $[K]$ &  $576,288$ $[Pa]$  & $1$   \\
    \hline
    \hline
\end{tabular}
\label{tab:ThrusterMach2ExpVarValues}
\end{table}

\rev{
\subsection{Aerospike Nozzle Flow: Three-Dimensional Perfectly Expanded Flow}
\label{sec:CaseSetup_3D}

We simulate three-dimensional, perfectly expanded aerospike nozzle flow to investigate the effect of the \Hybrid scheme's numerical dissipation on three-dimensional flow structures and regions with turbulence.
Initial and boundary conditions are identical to those of the perfectly expanded, two-dimensional nozzle flow (\cref{tab:ThrusterMach2NonExpVarValues}).
To reduce the computational cost of the simulation,  we truncate the computational domain to $x \in [-1.26, 10]$, $y \in [-3, 3]$, and the minimum grid spacing is increased to $\frac{1}{\Delta  s} = 32$.
The two-dimensional grid is extruded one length unit in the $z$ direction, spanning $z$ $ \in [0, 1]$.
The three-dimensional simulation is too computationally intensive to permit a domain-size study of the extrusion length; however, an extrusion length of one 
length unit spans the largest eddy scales evident in two dimensions (i.e., on the order of the spike-base width).
Periodic boundary conditions are applied to the spanwise ($z$) faces of the domain (\cref{fig:ASIC3D}).
All other conditions are unchanged as compared to the two-dimensional nozzle flow, and we use the CM05 blending model.
We follow the recommendations of Hennemann et al. \cite{Hennemann2021_JCP} and slightly increase $\alpha_{min}$ from $0$ to $0.0001$, which reduces the numerical oscillations in shock-adjacent regions.

\begin{figure}[ht]
    \centering
    \includegraphics[width=0.48\textwidth]{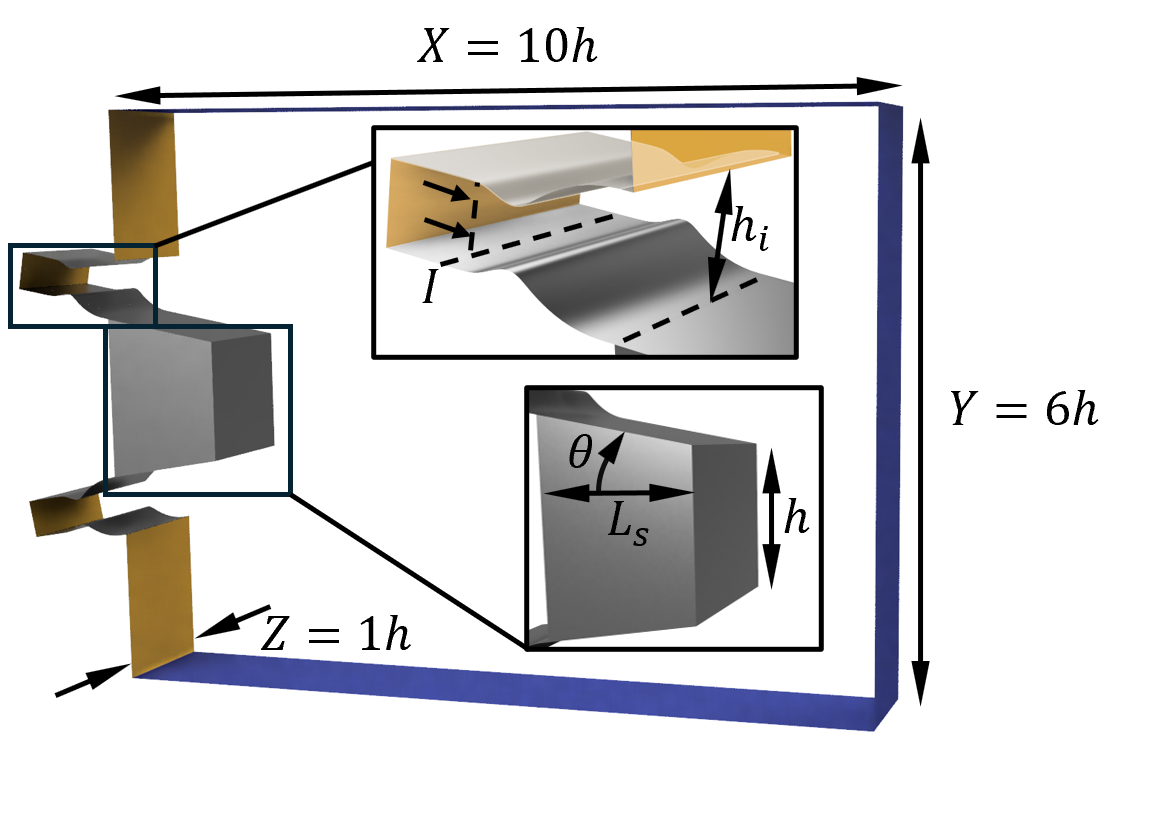}
    \caption{\rev{The computational domain of the three-dimensional aerospike case.}
    }
\label{fig:ASIC3D}
\end{figure}
}

\section{Results}
\label{sec:Results}

We assess the error convergence and computational efficiency of the \Hybrid scheme over the square cylinder wake.
We establish grid independence and investigate the effect of the \EV coefficient $C_m$ and the \FVSE blending coefficient $\alpha$ on aerospike nozzle flows using the perfectly expanded (2D) case.
Numerical stability in the hypersonic regime and in three-dimensional computations is demonstrated in the underexpanded (2D) case and perfectly expanded (3D) case, respectively.

\subsection{Square Cylinder Flow: Convergence and Computational Efficiency}
\label{subsec:GridStudy}

\rev{
The nominally incompressible flow over a square cylinder at a freestream Mach number $M = 0.1$ and Reynolds number $Re_f=100$ is characterized by a periodic vortex shedding and a von Karman vortex street in the far wake. 
In the time-averaged flow field, a recirculation region is observed behind the cylinder that reattaches at the centerline at the rear reattachment point, $x_u$ \cite{Klose2020_CF}.
We evaluate the convergence rate of the \Hybrid solver using this reattachment point.
To do so, we systematically refine the grid for $N = 3$ and determine the approximation error by comparison with the solution computed on a grid with $\frac{1}{\Delta s} = 64$.
}

\begin{figure}[htbp]
    \centering
        \begin{subfigure}[t]{0.51\textwidth}
        \includegraphics[width=0.9\linewidth, height=0.45\textwidth]{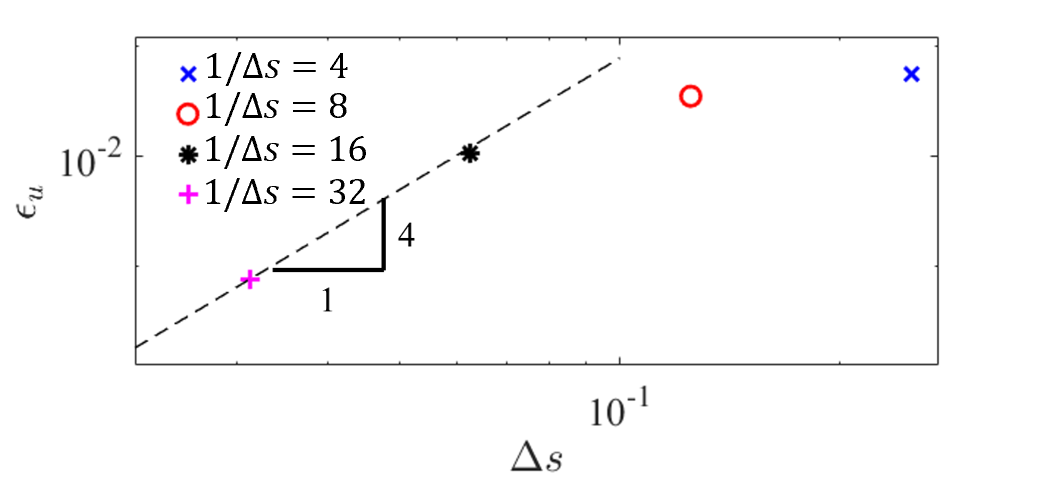}
        \caption{ }
        \label{fig:SqCylError}
    \end{subfigure}
    \begin{subfigure}[t]{0.47\textwidth}
        \includegraphics[width=1.0\linewidth, height=0.425\textwidth]{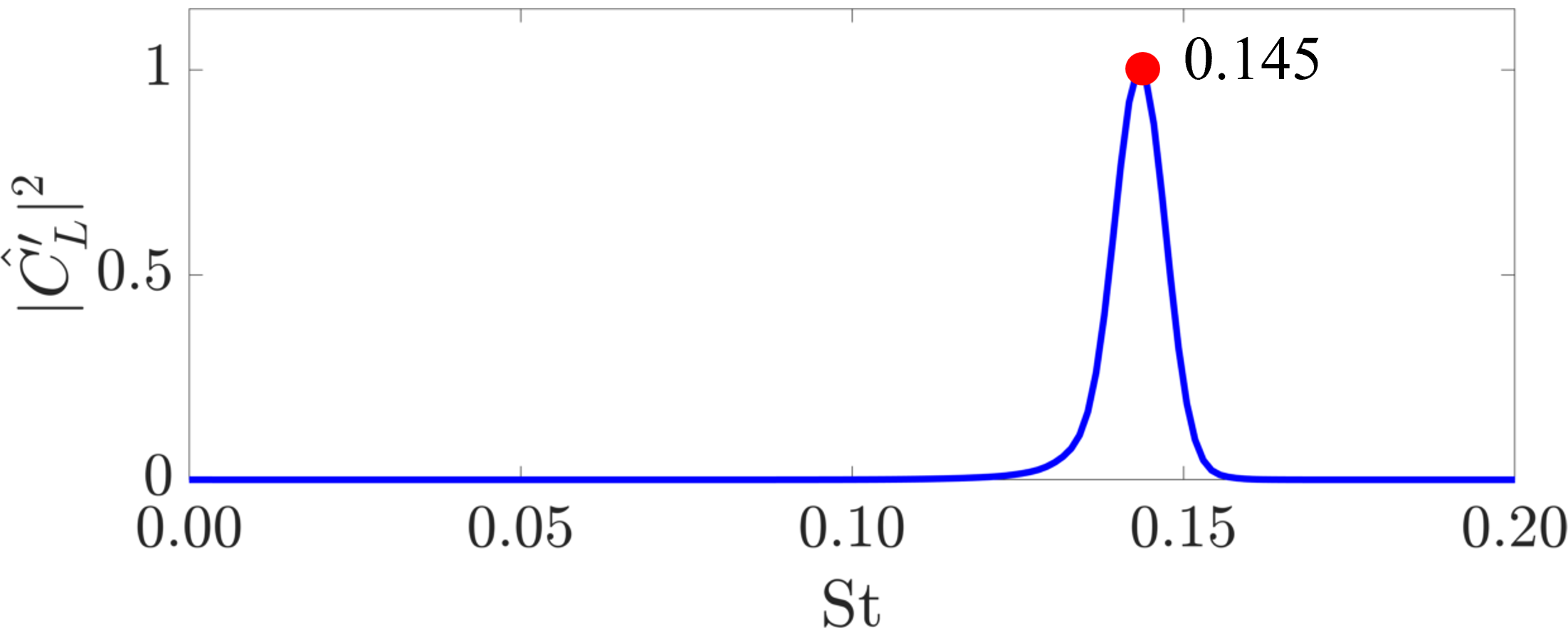}
        \caption{ }
        \label{fig:SqCylFreq}
    \end{subfigure}
    \caption{(a) The error in reattachment location and (b) the power spectrum of the square cylinder wake. 
    The $St = 0.145$ side-loading frequency is marked by a red dot.
    }
    \label{fig:SqCyl}
\end{figure}

The log-log plot of this error versus grid spacing, $\Delta s$, in \cref{fig:SqCylError} shows that the solution converges at the theoretical algebraic convergence rate of $N+1 = 4$ for smooth \DGSEM solutions \cite{cockburn2004_ECM, Kopriva2017_JSC}. 
From this result, we see that the hybridization of \DGSEM, \EV and \FVSE does not negatively impact the convergence rate of smooth flows, as also reported by Hennemann et al. \cite{Hennemann2021_JCP} for their \FVSE-based shock-capturing scheme.
Additionally, we compute the Strouhal number of the time-dependent lift (or side force) coefficient, $C_L$, to be $St = 0.145$ using the $\frac{1}{\Delta s} =32$ mesh (\cref{fig:SqCylFreq}).
This is in excellent agreement with the findings of Klose et al. \cite{Klose2020_CF} and indicates that the \Hybrid solver does not introduce dispersive error as compared to the split-form \DGSEM.

\rev{
The computational overhead of the \Hybrid solver as compared to the \DGSEM solver is determined from simulations with the $\frac{1}{\Delta s} = 32$ grid.  The \FVSE solution is computed in all elements.
For this case, the \Hybrid scheme requires $26.6\%$ more compute time than the solver without shock-capturing and $25.4\%$ more than the \EV scheme.
We note that the computational cost can be significantly reduced by  evaluating the \FVSE solution only in elements where $\alpha$ exceeds a prescribed threshold (e.g., $\alpha > 0.0001$).
}

% \FloatBarrier

\subsection{Aerospike Nozzle Flow: Two-Dimensional Perfectly Expanded Flow}
\rev{
\label{sec:Results_Mach2NonExpanding}

\subsubsection{Grid Independence Study}

To evaluate the convergence behavior of the \Hybrid solver for simulations of aerospike nozzle flows that are characterized by unsteady wakes with strong shocks, we compute the perfectly expanded Mach $2$ flow on five meshes that are systematically refined with minimum grid spacings of: $\frac{1}{\Delta s} = 8, 16, 32, 64, 128$. 
The convergence metrics are the time-averaged rear reattachment point, $x_u$, and the time-averaged pressure integrated over the base of the spike, $\overline{p}_b$.
The error in the mean reattachment point location measures statistical convergence, and the integrated base pressure is commonly used in grid independence studies of steady-state aerospike nozzle flows \cite{Hagemann1998_AIAAConf, Nair2017_JTS, Nair2019_PPR}.
We treat the solution computed on the $\frac{1}{\Delta s} = 128$ grid as a reference solution to determine the errors on coarser grids.
We approximate the solution with an $N = 3$ polynomial in all elements.

\begin{figure}[H]
    \centering
    \begin{subfigure}[b]{0.45\textwidth}
        \centering
        \includegraphics[width=0.8\textwidth]{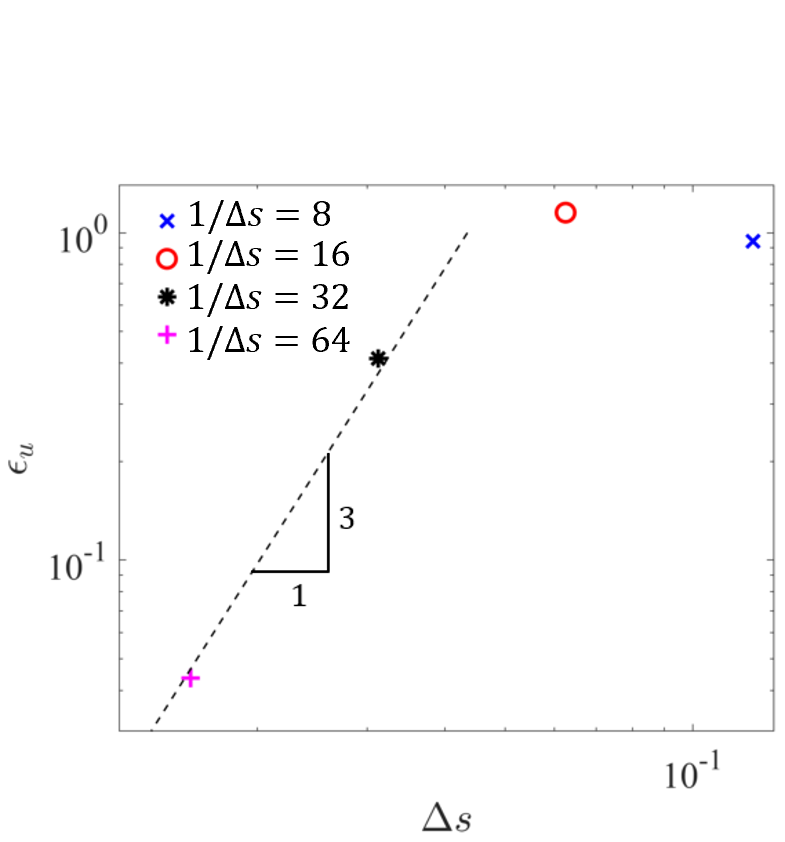}
        \caption{ }
        \label{fig:StagUErr}        
    \end{subfigure}
    \begin{subfigure}[b]{0.45\textwidth}
        \centering
        \includegraphics[width=0.8\textwidth]{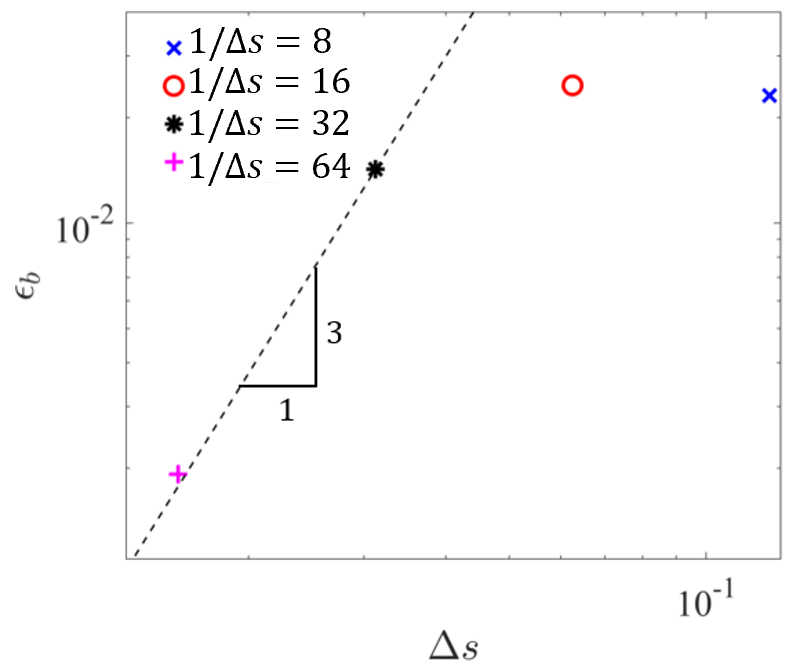}
        \caption{ }
        \label{fig:IntBasePErr}        
    \end{subfigure}
    \hfill
    \caption{The error of (a) the mean jet reattachment location, $\epsilon_u$,  and (b) the mean nondimensional base pressure, $\epsilon_b$, plotted against grid spacing.
    }
    \label{fig:GridStudyErr}
\end{figure}

Log-log plots of the error vs grid spacing for $x_u$ and $\overline{p}_b$ (\cref{fig:GridStudyErr}) yield slopes of $3.24$ and $2.88$, respectively.
The logarithmic error decay indicates that the solutions computed on the grids with $\frac{1}{\Delta s} = 64$ and $\frac{1}{\Delta s} = 32$ have achieved grid independence.
The convergence rate is one order of magnitude less compared to the rate for smooth \DGSEM solutions.
We attribute this to a well-known reduction of numerical accuracy to first order in the capturing of shock discontinuities. 
The approximate $\mathcal{O}(N)$ convergence rate of the global solution is greater than the local $\mathcal{O}(1)$ accuracy close to shocks, as also  reported by Peck et al. \cite{Peck2024_AIAA} for  \FVSE schemes.
All two-dimensional computations in the following sections are performed on grids with $\frac{1}{\Delta s} = 64$ with a grid size of $337{,}966$ elements.

% and $5{,}407{,}456$ degrees of freedom while the three-dimensional grid contains $1{,}571{,}520$ elements and $100{,}577{,}280$ degrees of freedom.
% \rev{
% % Three-dimensional computations are conducted with on grids with $\frac{1}{\Delta s} = 64$  over a truncated computational domain, described in \cref{sec:CaseSetup_3D}. 

% }

\subsubsection{Effect of Blending Model}

\rev{

% We compare three blending models for the hybrid scheme (see \cref{sec:HybridScheme}), including a constant global blending (A01), and two shock indicator–based blends with $C_m=0.1$ (CM01) and $C_m=0.5$ (CM05).
% These are compared to a reference entropy viscosity scheme with a spectral filter (EVF) that is developed by Chaudhuri et al. \cite{Chaudhuri2017_JCP} and employed in short-time aerospike nozzle computations by Chai et al. \cite{Chai2021_AIAAConf}.
We compare the blending models as described in \cref{sec:HybridScheme}, including a constant global blending (A01), and two shock indicator–based blends with $C_m=0.1$ (CM01) and $C_m=0.5$ (CM05), and find that the solution remains bounded in all models for at least $500$ convective time units.
In contrast, the EVF scheme \cite{Chaudhuri2017_JCP} uses an exponential filter instead of \FVSE and is unstable after $23$ convective time units.
Numerical schlieren images in \cref{fig:Mach2NonExp_FVSESlrn} show that the A01 model is overly dissipative and suppresses  waves and vortices observed in the CM01 and CM05 cases.
In fact, the A01 case is qualitatively similar to RANS simulations of comparable aerospike nozzle flows \cite{Nair2017_JTS, Nair2019_JSR, Nair2019_PPR, Pyle2023_AIAAConf}, in which the turbulent Reynolds stresses are modeled using eddy-viscosity models.
In the CM01 and CM05 cases, the $\alpha$ contours in \cref{fig:Mach2NonExp_FVSEAlpha} correspond to shocks observed in the numeric schlieren images \cref{fig:Mach2NonExp_FVSESlrn}.
The $\alpha$ contours (\cref{fig:Mach2NonExp_FVSEAlpha}) appear pixelated because $\alpha$ is constant over a given element [\cref{eq:AlphaM}], rather than a continuous $N^{th}$ order polynomial.
In the EVF case, a checkerboard pattern forms near strong shocks where the exponential filter is applied (\cref{fig:EVFSnapshot}), which is consistent with aerospike nozzle simulations reported by Chai et al. \cite{Chai2021_AIAAConf}.
The checkerboard pattern leads to  significant solution discontinuities at element interfaces that eventually lead to negative densities and numerical instability.

\begin{figure}[htbp]
    \begin{subfigure}[t]{0.49\textwidth}
        \includegraphics[width=1.0\textwidth]{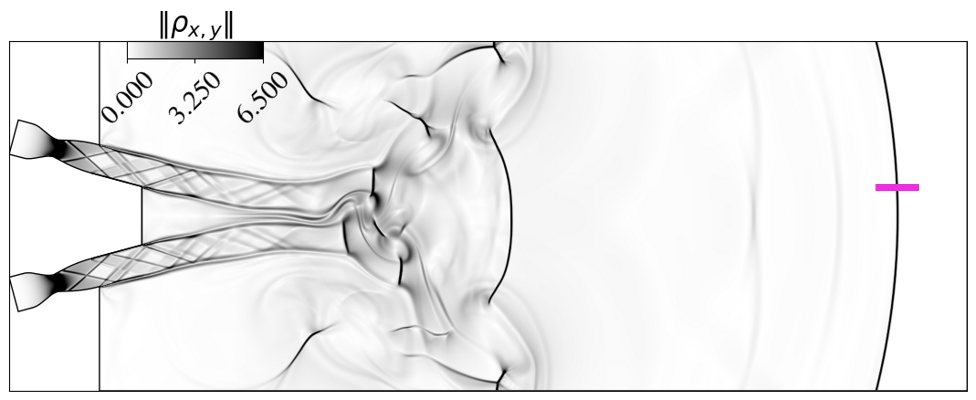}
        \includegraphics[width=1.0\textwidth]{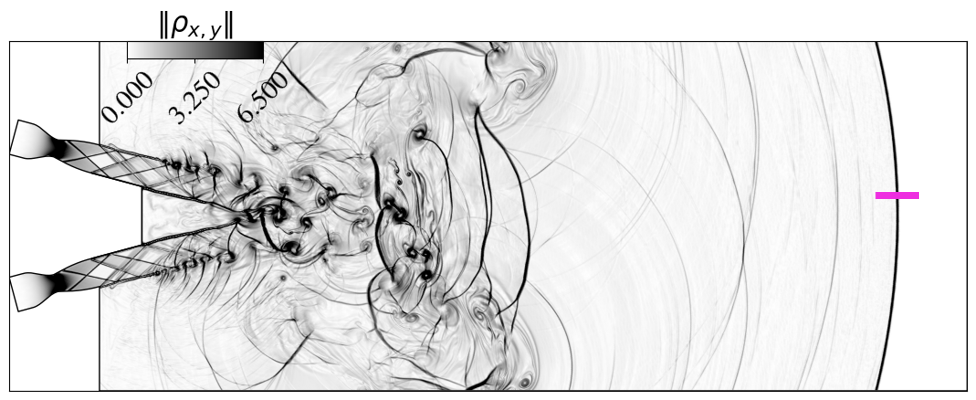}
        \includegraphics[width=1.0\textwidth]{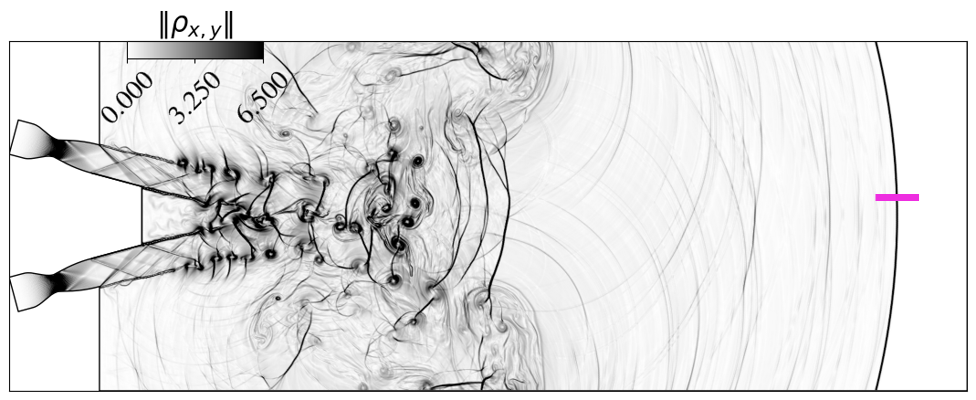}
        \caption{ }
        \label{fig:Mach2NonExp_FVSESlrn}
    \end{subfigure}
    \begin{subfigure}[t]{0.49\textwidth}
        \includegraphics[width=1.0\textwidth]{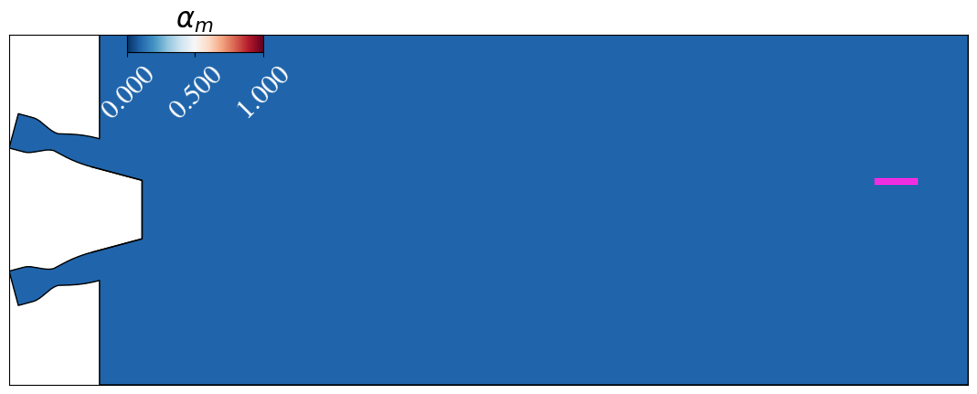}
        \includegraphics[width=1.0\textwidth]{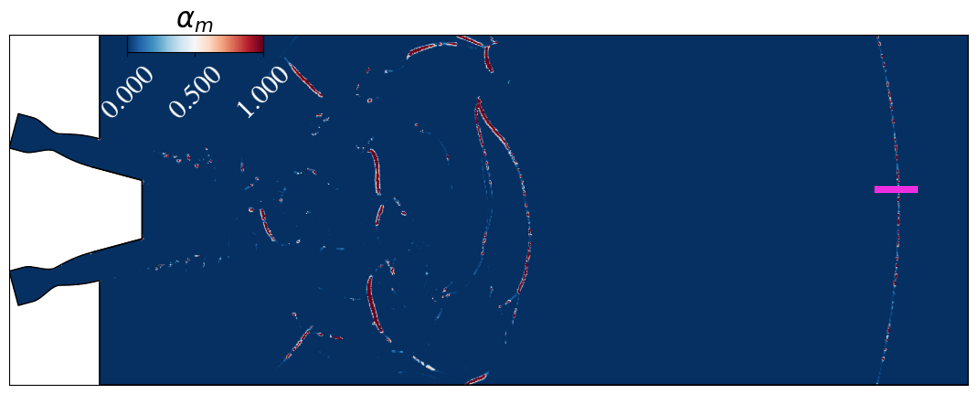}
        \includegraphics[width=1.0\textwidth]{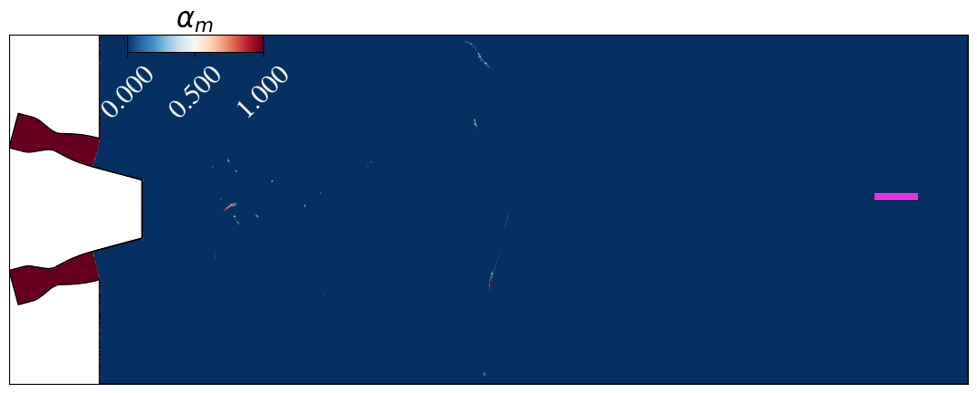}
        \caption{ }
        \label{fig:Mach2NonExp_FVSEAlpha}
    \end{subfigure}
    \caption{\rev{Contours of (a) numerical schlieren images and (b) the \FVSE blending coefficient $\alpha$. 
    From top to bottom, the models are: A01, CM01, and CM05.}
    }
    \label{fig:Mach2NonExp_FVSESlrnAlpha}
\end{figure}

\begin{figure}[H]
    \centering
    \includegraphics[width=0.925\linewidth]{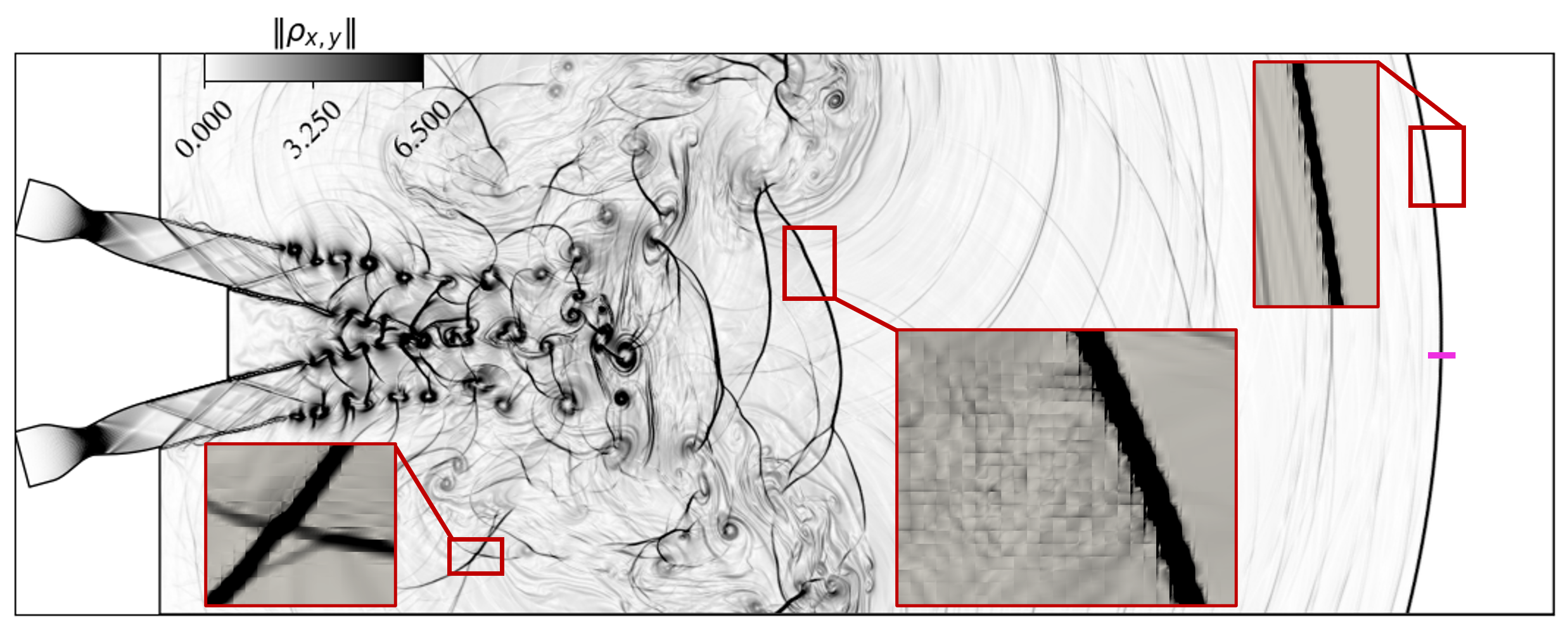}
    \caption{\rev{
    Numerical schlieren image of the reference EVF case. 
    The pixelated effect indicates regions where a spectral filter is applied.}
    }
    \label{fig:EVFSnapshot}
\end{figure}

% Increasing the $C_m$ coefficient for case CM05 as compared to CM01 
% decreases the shock indicator $\Psi$, and thus the \FVSE blending coefficient $\alpha$, according to \cref{eq:PsiEq}.
The number of elements in which $\mu_h > 0$ in the CM05 case is within $1.2 \%$ of that obtained with the CM01 model. 
However, a comparison of the contour plots $\Psi$ in \cref{fig:PsiContourComp} shows that, in general, $\Psi$ has a smaller magnitude at shock locations for the CM05 model than the CM01 model. 
Increasing $C_m$, and therefore $\mu_{max}$, in the CM05 case indirectly reduces the \FVSE blending coefficient $\alpha$ via the shock indicator [\cref{eq:PsiEq}].
The decrease in $\alpha$ leads to the \EV scheme increasingly carrying the high-resolution shock-capturing, whereas the dissipative \FVSE blending is significant only in elements where $\mu_h \approx \mu_{max}$.
Comparing the $\alpha$ contours in \cref{fig:Mach2NonExp_FVSEAlpha} between the CM01 and CM05 models, it is apparent that significantly fewer elements satisfy $\alpha > 0$ for CM05.  
In fact, the blending coefficient $\alpha$ is non-zero in  approximately three times more elements in the CM01 case as compared to the CM05 case.
Averaging $\alpha$ over the computational domain  yields $\overline{\alpha}=0.015$ and $0.007$ in the CM01 and CM05 cases, respectively.
Both of these metrics provide evidence of an increased \FVSE-based dissipation in the CM01 case.
The effect this dissipation has on the flow is noticeable.
Both CM01 and CM05 exhibit shear-layer instabilities along the outer shear layer that deflect the jet cores and generate internal compression waves (\cref{fig:Mach2NonExp_FVSESlrn}). 
In the CM05 case, these internal waves perturb the inner shear layers and generate instabilities; 
however, no perturbations are present along the inner shear layers in the CM01.

\begin{figure}[htbp]
    \begin{subfigure}[t]{0.49\textwidth}
        \includegraphics[width=1.0\textwidth]{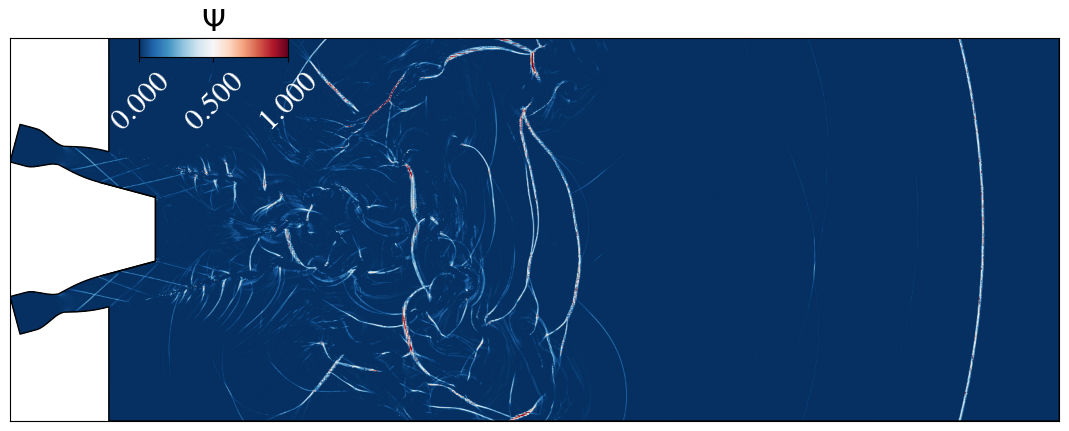}
        \caption{ }
        \label{fig:CM01Psi}
    \end{subfigure}
    \begin{subfigure}[t]{0.49\textwidth}
        \includegraphics[width=1.0\textwidth]{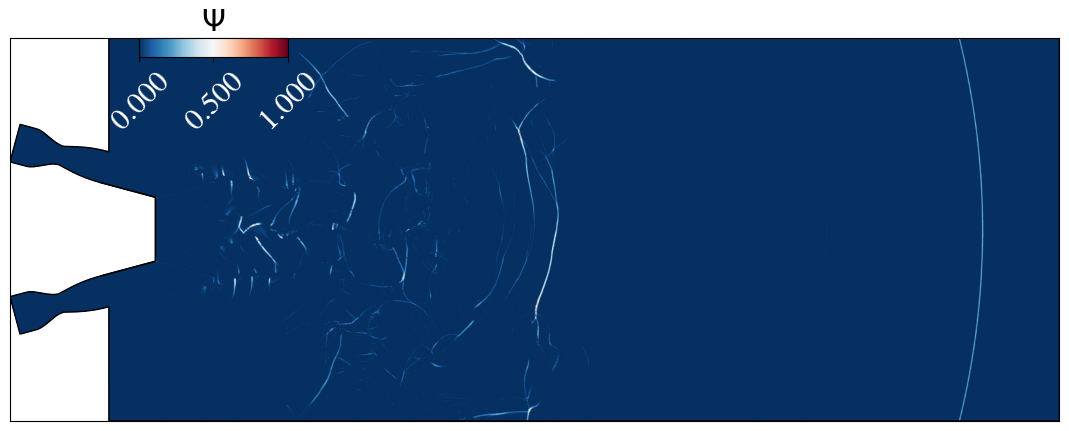}
        \caption{ }
        \label{fig:CM05Psi}
    \end{subfigure}
    \caption{\rev{Contours of the shock indicator, $\Psi$, using (a) the CM01, and (b) the CM05 model.}
    }
    \label{fig:PsiContourComp}
\end{figure}

We extract the density profile along the pink line in \cref{fig:Mach2NonExp_FVSESlrnAlpha}, i.e., across the initial shock that advects downstream, and compare the effect of the blending models on the capturing of this shock. 
We interpolate the density profiles with the element's polynomial basis and plot them in \cref{fig:RhoAcrossShock}. 
The vertical dashed lines demarcate the boundaries of the shock-containing element.
In the A01 and CM01 cases, $\rho$ decreases monotonically across the shock.
The monotonicity and positivity-preserving properties of the \LLF scheme is guaranteed in the limit $\alpha \to 1$.
However, we see in the A01 case that blending coefficients as low as $\alpha = 0.1$  produce monotone solutions as well.
A slight undershoot is observed in the CM05 case, in which $\alpha \approx 0$ across the shock.
Both overshoots and undershoots are noticeable in the EVF case, demonstrating that the hybrid \EV-\FVSE scheme is more effective at positivity-preserving shock-capturing than the EVF scheme.

\begin{figure}[htbp]
    \centering
    \includegraphics[width=0.5\linewidth]{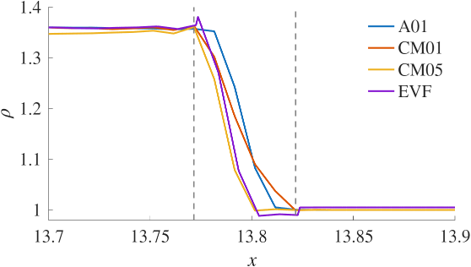}
    \caption{\rev{
    Density $\rho$ extracted across a shock. The extraction location is indicated by the pink line in \cref{fig:Mach2NonExp_FVSESlrn}.
    }}
    \label{fig:RhoAcrossShock}
\end{figure}

\begin{figure}[htbp]
    \centering
    \begin{subfigure}[t]{0.45\textwidth}
        \includegraphics[width=1.0\textwidth]{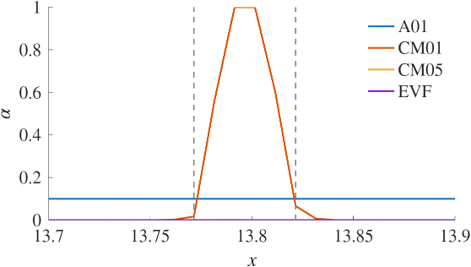}
        \caption{ }
        \label{fig:AlphaAcrossShock}
    \end{subfigure}
    \begin{subfigure}[t]{0.45\textwidth}
        \includegraphics[width=1.0\textwidth]{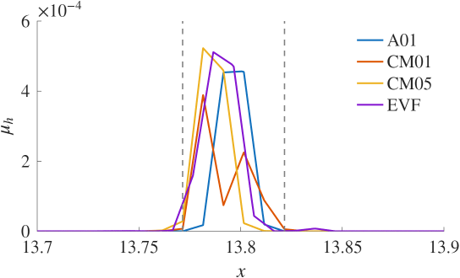}
        \caption{ }
        \label{fig:MuAcrossShock}
    \end{subfigure}
\caption{\rev{
    (a) The entropy viscosity $\mu_h$ and (b) blending coefficient $\alpha$ across a shock in the A01, CM01, and CM05 cases. The extraction location is indicated by the pink line in \cref{fig:Mach2NonExp_FVSESlrn}.
}}
\label{fig:FVSEAcrossShock}
\end{figure}

A monotonic numerical solution in  elements containing shocks does not, by itself, ensure numerical stability. 
For example, the blending coefficient 
$\alpha$ reaches unity within the shocked element in the CM01 case (\cref{fig:AlphaAcrossShock}), yet oscillations are present in the upstream element (\cref{fig:RhoAcrossShock}).
These are caused by the element-coupling via the two-sided \LLF inter-element flux [\cref{eq:LxFFlux}].
Through this coupling, the sharp gradient within the shock-containing element affects the solution in adjacent elements and can induce numerical oscillations.
These oscillations may produce negative densities and cause the simulation to crash, despite a monotone solution over the shock.

% Non-monotonic solutions are observed only in the upstream element and not in the downstream element because the shock location as identified  by the 
% maxima in the $\mu_h$ plots in \cref{fig:MuAcrossShock} is slightly biased toward the upstream element boundary.
Hennemann et al. \cite{Hennemann2021_JCP} suggest a general treatment to damp oscillations in shock-adjacent elements, which specifies  a non-zero, global minimum for the blending coefficient $\alpha_{min} > 0$. 
For their \FVSE scheme, they report excellent stability characteristics with $\alpha_{min} = 0.0001$, which yields a numerical dissipation  at all solution locations. 
The global dissipation may also serve as an implicit subgrid-scale viscosity model, similar to low-pass filters employed in implicit LES (ILES) \cite{Margolin2006_JT}.
Another approach is to raise $\alpha$ in shock-adjacent elements. 
Peck et al. \cite{Peck2024_AIAA} uses this approach and report an increase in stability at the cost of increased overhead for the computation of $\alpha$.

}
%\clearpage
\FloatBarrier
}

\subsection{Aerospike Nozzle Flow: Two-Dimensional Underexpanded Flow}
\rev{
 \label{sec:3_Mach5Expanding}

To assess the robustness of the hybrid shock-capturing scheme in the low-hypersonic regime, we simulate flow expanding from a thruster exit Mach number of $M=2$ to $M=5$ over the aerospike nozzle using the CM05 blending model.
Contours of the Mach number and $\alpha$ are plotted at times $t_1=2.25$, $t_2=6.25$, $t_3=15$, and $t_4=100$ (\cref{fig:ThrusterMach5Exp_MachAlphaMax}).
At time $t_1$, a convex, initial shock front develops that propagates downstream.
By time $t_2$, a secondary shock front forms that connects the reattachment shocks to the outer shear layer, and an upstream-propagating wave is observed in the nascent recirculation region.
A non-linear flow instability at the reattachment point at $t_3$ leads to the formation of a train of vortices.
A quasi-steady state is achieved by $t_4$, and a shock-free vortex street is observed along the centerline.

\begin{figure}[ht]
    \begin{subfigure}[t]{0.48\textwidth}
        \includegraphics[width=1.0\textwidth, height=0.4\textwidth]{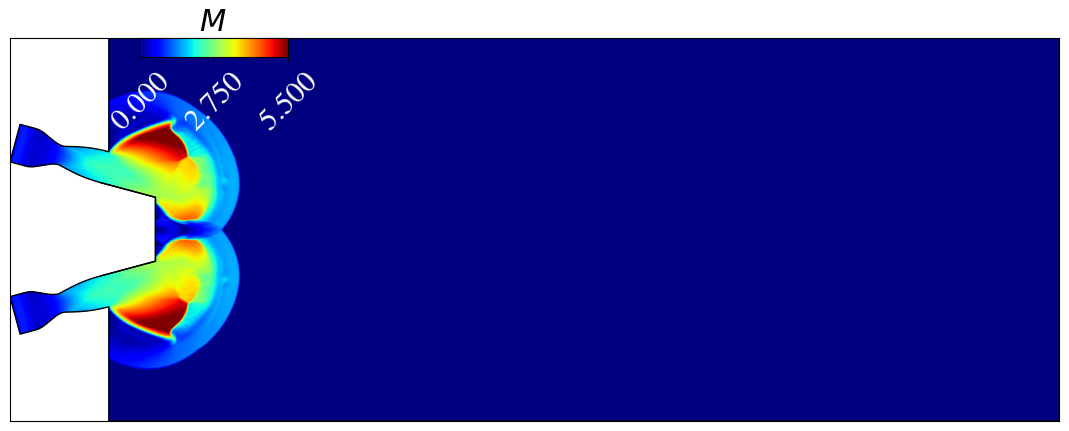}
        \includegraphics[width=1.0\textwidth, height=0.4\textwidth]{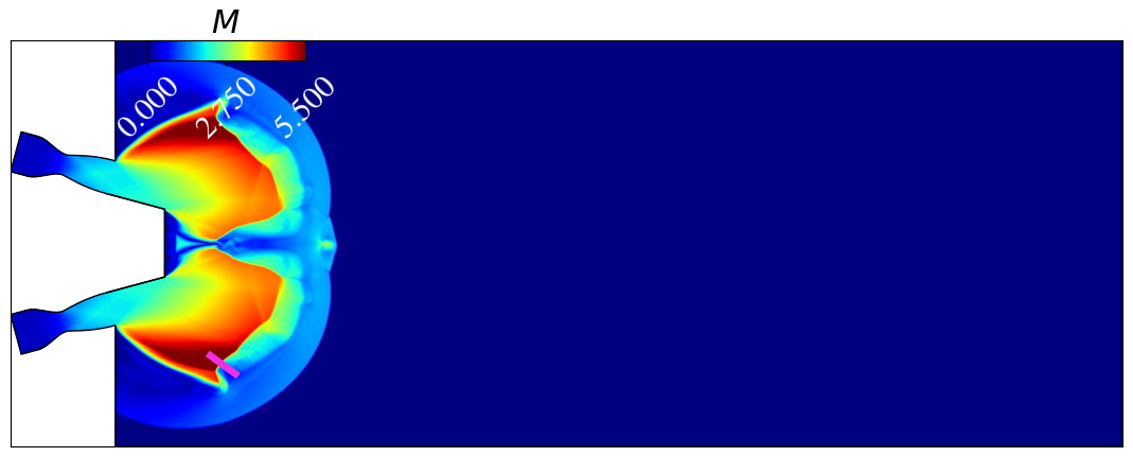}
        \includegraphics[width=1.0\textwidth, height=0.4\textwidth]{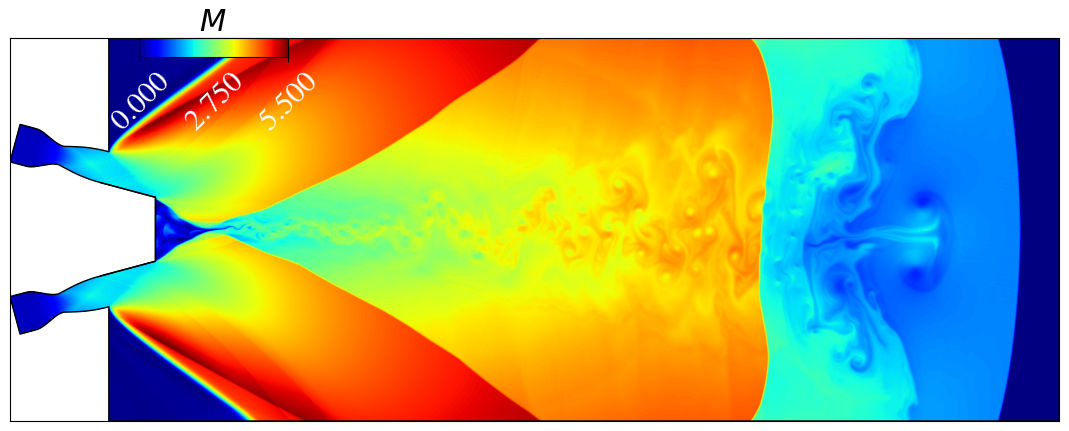}
        \includegraphics[width=1.0\textwidth, height=0.4\textwidth]
        {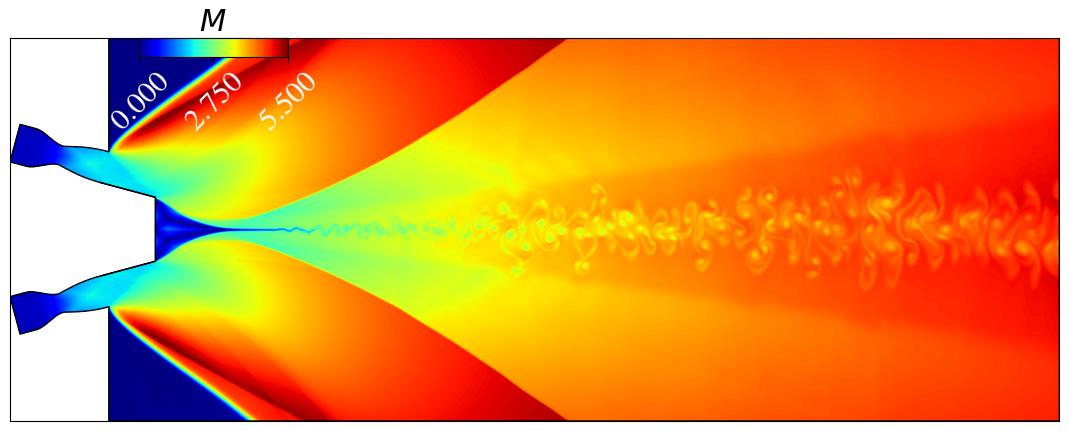}
        \caption{ }
        \label{fig:ThrusterMach5ExpMach}
    \end{subfigure}
    \begin{subfigure}[t]{0.48\textwidth}
        \includegraphics[width=1.0\textwidth, height=0.4\textwidth]{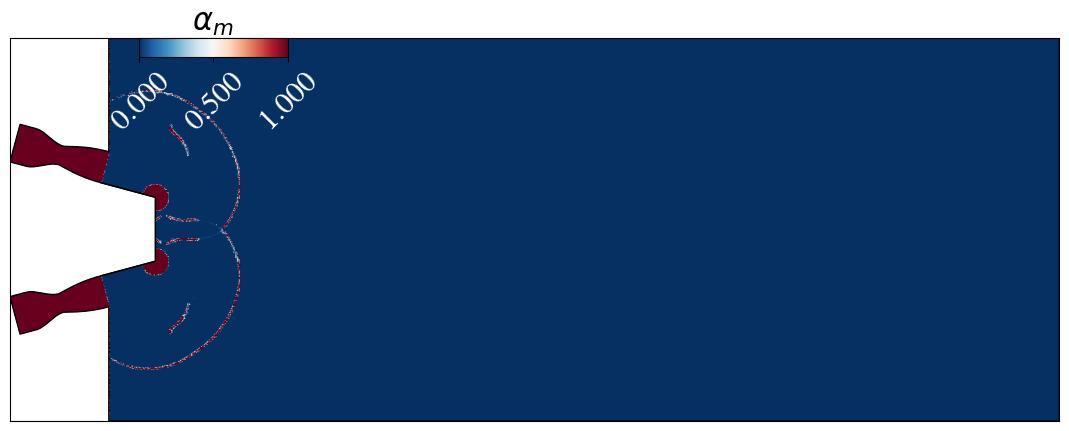}
        \includegraphics[width=1.0\textwidth, height=0.4\textwidth]{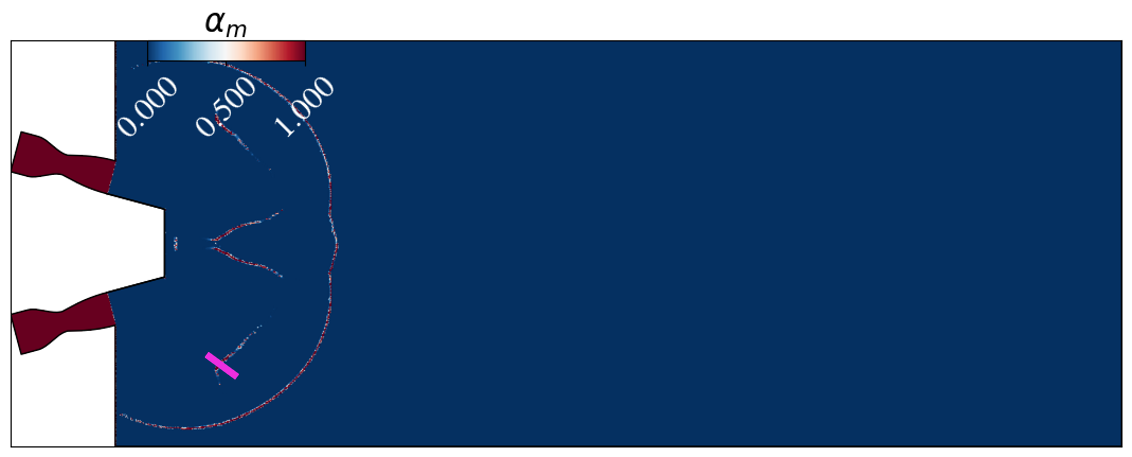}
        \includegraphics[width=1.0\textwidth, height=0.4\textwidth]{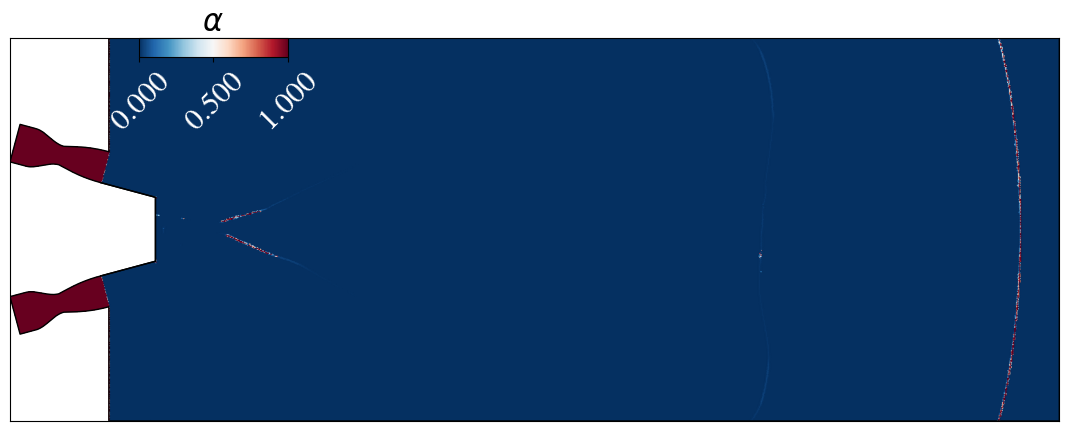}
        \includegraphics[width=1.0\textwidth, height=0.4\textwidth]{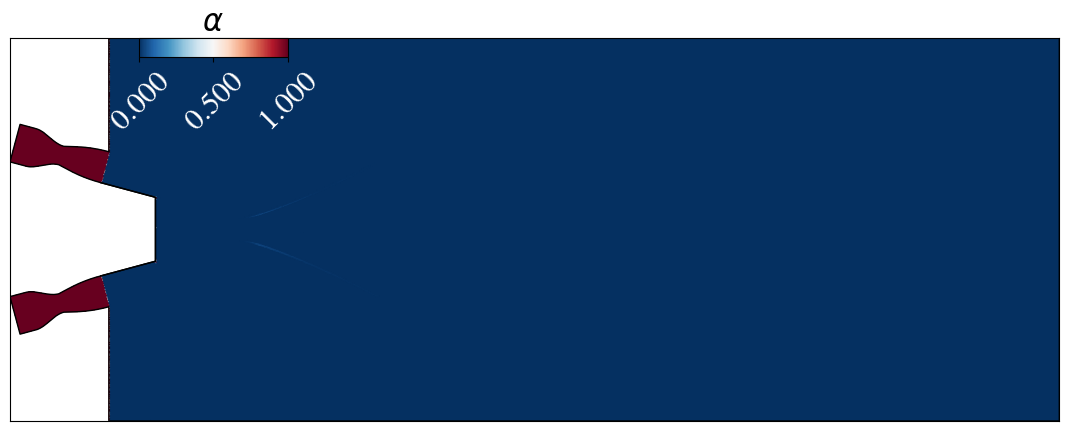}
        \caption{ }
        \label{fig:ThrusterMach5ExpAlphaMax}
    \end{subfigure}
    \caption{\rev{
    Snapshots of (a) Mach contours and (b) $\alpha$ of the underexpanded case at $t_1=2.25$, $t_2=6.25$, $t_3=15$, and $t_4=100$.}
    }
    \label{fig:ThrusterMach5Exp_MachAlphaMax}
\end{figure}

Profiles of Mach number, pressure, $\mu_h$, and $\alpha$ are extracted along  the pink line in \cref{fig:ThrusterMach5Exp_MachAlphaMax}, i.e, across the oblique secondary shock front at $t_2$, and plotted in \cref{fig:FVSEAcrossShock_M5,fig:FlowAcrossShock_M5}.The blending coefficient 
$\alpha$ reaches unity across the shock (\cref{fig:AlphaAcrossShock_M5}), meaning $\mu_h = \mu_{max}$.
However, comparing \cref{fig:MuAcrossShock_M5} and \cref{fig:MuAcrossShock}, we observe that $\mu_h$ is approximately three times lower in the underexpanded case than in the perfectly expanded case.
The drop in density and temperature associated with the jet's expansion decreases $\mu_{max}$ according to \cref{eq:MuAVMax}, which increases the \FVSE blending coefficient [\cref{eq:PsiEq,eq:BlendingFunction_AlphaTilde}].
As a result, the dissipative \FVSE solution is blended more liberally in underexpanded aerospike nozzle flows where strong shocks, i.e., high upstream Mach numbers (\cref{fig:MAcrossShock_M5}), are present.
The increased dissipation produces a monotone solution across the shock and mitigates undershoots in the already-low density (\cref{fig:RAcrossShock_M5}) that may lead to negative values.
This adaptive behavior is a result of using the \EV-\FVSE hybridization.
Other common sensors used in subcell shock-capturing methods, such as the Persson indicator \cite{Persson2006_AIAAConf, Hennemann2021_JCP}, are not inherently more aggressive in low-density regions.

\begin{figure}[htbp]
    \centering
    \begin{subfigure}[t]{0.49\textwidth}
        \includegraphics[width=1.0\textwidth]{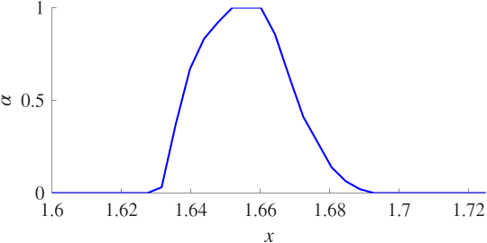}
        \caption{ }
        \label{fig:AlphaAcrossShock_M5}
    \end{subfigure}
    \begin{subfigure}[t]{0.49\textwidth}
        \includegraphics[width=1.0\textwidth]{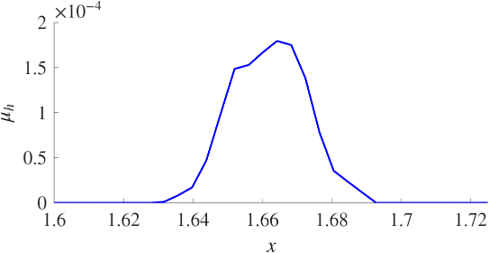}
        \caption{ }
        \label{fig:MuAcrossShock_M5}
    \end{subfigure}
\caption{\rev{
    (a) Blending coefficient $\alpha$ and (b) entropy viscosity $\mu_h$ extracted across a shock at $t_2$ in the underexpanded case.
}}
\label{fig:FVSEAcrossShock_M5}
\end{figure}

\begin{figure}[htbp]
    \centering
    \begin{subfigure}[t]{0.49\textwidth}
        \includegraphics[width=1.0\textwidth]{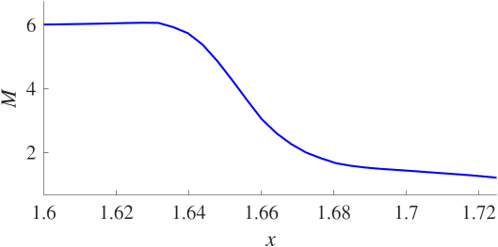}
        \caption{ }
        \label{fig:MAcrossShock_M5}
    \end{subfigure}
    \begin{subfigure}[t]{0.49\textwidth}
        \includegraphics[width=1.0\textwidth]{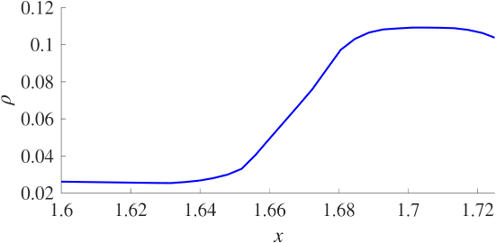}
        \caption{ }
        \label{fig:RAcrossShock_M5}
    \end{subfigure}
\caption{\rev{
    (a) The Mach number and (b) density extracted across a shock at $t_2$ in the underexpanded case.
}}
\label{fig:FlowAcrossShock_M5}
\end{figure}

Profiles of density, $\mu_h$, and $\alpha$ versus the $x$-coordinate along the symmetry line are plotted in \cref{fig:RhoAcrossCenterline,fig:MuAcrossCenterline,fig:AlphaAcrossCenterline}. 
The blending coefficient $\alpha$ increases to unity at several locations (\cref{fig:AlphaAcrossCenterline}) that align with the sharp density gradients demarcating the leading and secondary shock fronts, jet reattachment point, and upstream-propagating recirculation shock (\cref{fig:RhoAcrossCenterline}).
As in the perfectly expanded Mach $2$ case, the solution is monotonic across these shocks, but density fluctuations appear in the shock-adjacent elements.
In this case, we attribute the density oscillations to physical fluctuations associated with flow structures rather than numerical oscillations.
For example, the density fluctuations over $x \in [3, 11]$ at $t_3$ (\cref{fig:RhoAcrossCenterline}) correspond to the vortices generated by the oscillatory reattachment point (\cref{fig:ThrusterMach5ExpMach}). 
% The Mach number in these vortices is greater than one, i.e., supersonic flow.
At $t_4$, the blending coefficient, $\alpha$, and entropy viscosity, $\mu_h$, are  both approximately zero along the centerline, suggesting that the unsteady oscillations at the reattachment point and the resulting vortex street are not attenuated by the shock-capturing scheme.  
A further analysis of the flow physics is beyond the scope of this paper, and we reserve a more detailed discussion for a future manuscript.

\begin{figure}[htbp]
    \centering
    \includegraphics[width=0.5\linewidth]{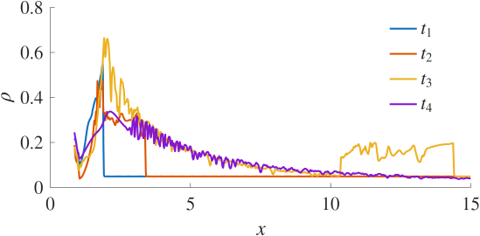}
    \caption{\rev{
    Density $\rho$ extracted across the centerline of the underexpanded case at $t_1=2.25$, $t_2=6.25$, $t_3=15$, and $t_4=100$.
    }}
    \label{fig:RhoAcrossCenterline}
\end{figure}

\begin{figure}[htbp]
    \centering
    \begin{subfigure}[t]{0.49\textwidth}
        \includegraphics[width=1.0\textwidth]{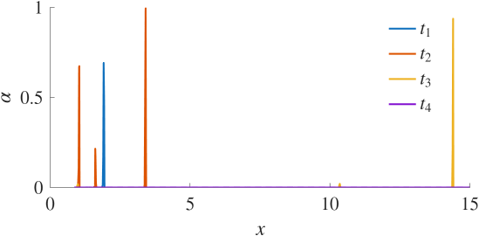}
        \caption{ }
        \label{fig:AlphaAcrossCenterline}
    \end{subfigure}
    \begin{subfigure}[t]{0.49\textwidth}
        \includegraphics[width=1.0\textwidth]{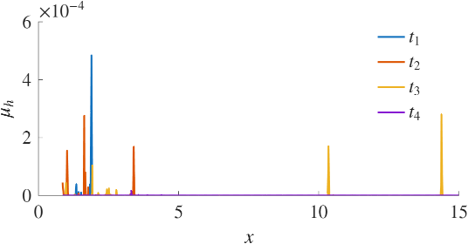}
        \caption{ }
        \label{fig:MuAcrossCenterline}
    \end{subfigure}
\caption{\rev{
     The (a) blending coefficient $\alpha$ and (b) entropy viscosity $\mu_h$ extracted across the centerline of the underexpanded case at $t_1=2.25$, $t_2=6.25$, $t_3=15$, and $t_4=100$.
}}
\label{fig:FVSEAcrossCenterline}
\end{figure}

\FloatBarrier
 }

\rev{
\subsection{Three-Dimensional Perfectly Expanded Flow}
\label{sec:3DResults}

To assess the \Hybrid solver in a three-dimensional environment, we simulate a perfectly expanded Mach $2$ flow over an aerospike nozzle (\cref{fig:ASIC3D}).
An inspection of the  iso-contours of the $Q$-criterion (\cref{fig:QCriterion3D}) and vorticity magnitude (\cref{fig:VortMag3D}) overlaid onto two-dimensional schlieren images of the $z = 0.5$ plane at $t = 2.25, 6.25, 15$, and $100$, show that
these results are in good qualitative agreement with the two-dimensional, perfectly expanded Mach $2$ test case (\cref{fig:Mach2NonExp_FVSESlrnAlpha}).
Startup vortices develop at the spike tip at  $t = 6.25$ and a wave develops in the recirculation region between the jets that connects these vortices. 
As the jets merge and collide, this wave propagates upstream and incites three-dimensional instabilities at the centerline. 
% We will discuss this instability in detail in a future manuscript and focus on the assessment of the \Hybrid scheme  for simulation of the aerospike nozzle flow here.  
At later times, the flow approaches a quasi-steady state and oscillations at the reattachment point are observed.
As in the two-dimensional flow, these oscillations lead to a  shedding of vortices in the upstream direction into the subsonic recirculation region and downstream into the oscillatory wake. 
Clearly, the two-dimensional computations are appropriate for cost-effective studies of resolution requirements and modeling parameters, as well as investigations into two-dimensional instabilities.
However, three-dimensional simulations are necessary to capture three-dimensional flow structures and instabilities in the shear layers and the vortex street along the wake centerline.
A detailed investigation into three-dimensional flow physics of the aerospike nozzle wake is outside the scope of this manuscript.
% but it is apparent that three-dimensional structures and the related unsteady dynamics are not suppressed.

\begin{figure}[htbp]
    \begin{subfigure}[t]{0.45\textwidth}
        \includegraphics[width=1.0\textwidth]{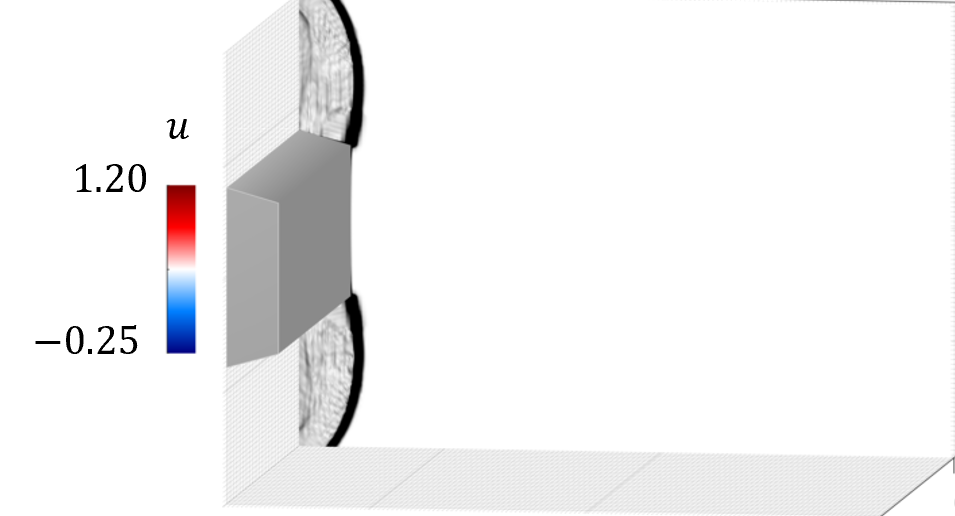}
        \caption{ }
        \label{fig:QCriterion3D1}
    \end{subfigure}
    \hfill
    \begin{subfigure}[t]{0.45\textwidth}
        \includegraphics[width=1.0\textwidth]{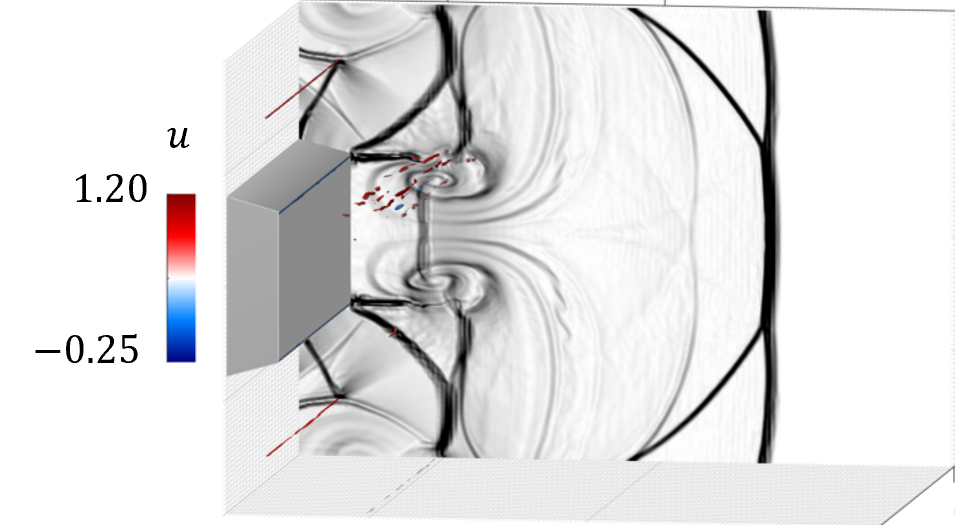}
        \caption{ }
        \label{fig:QCriterion3D2}
    \end{subfigure}
    \hfill
    \begin{subfigure}[t]{0.45\textwidth}
        \includegraphics[width=1.0\textwidth]{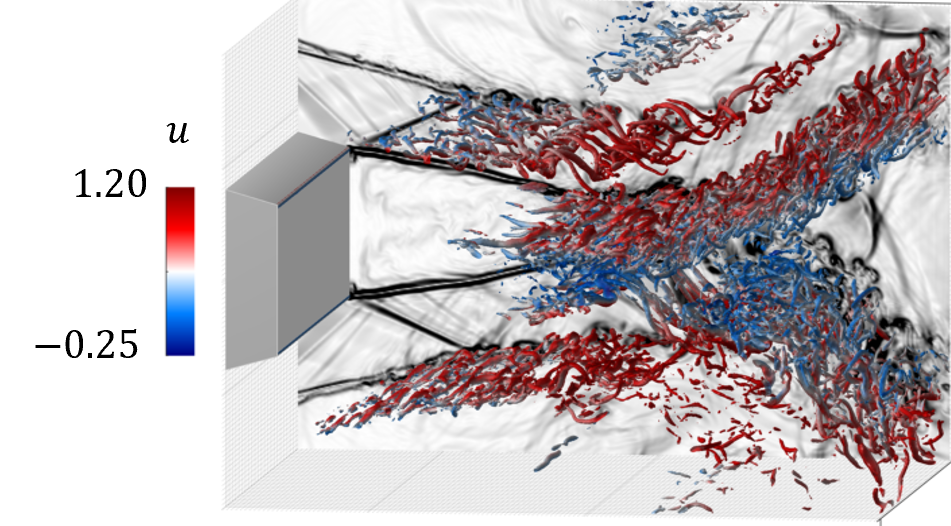}
        \caption{ }
        \label{fig:QCriterion3D3}
    \end{subfigure}
    \hfill 
    \begin{subfigure}[t]{0.45\textwidth}
        \includegraphics[width=1.0\textwidth]{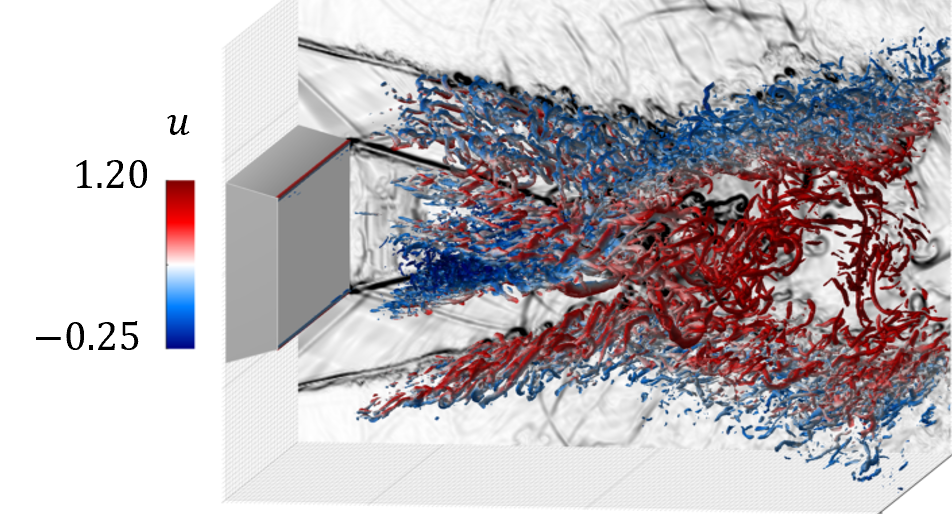}
        \caption{ }
        \label{fig:QCriterion3D4}
    \end{subfigure}
    \hfill
    \caption{\rev{Iso-contours of the $Q$-criterion at $Q = 40$ at times (a) $t = 2.25$, (b) $t = 6.25$, (c) $t = 15$, and (d) $t = 100$.}
    }
    \label{fig:QCriterion3D}
\end{figure}

\begin{figure}[htbp]
    \begin{subfigure}[t]{0.45\textwidth}
        \includegraphics[width=1.0\textwidth]{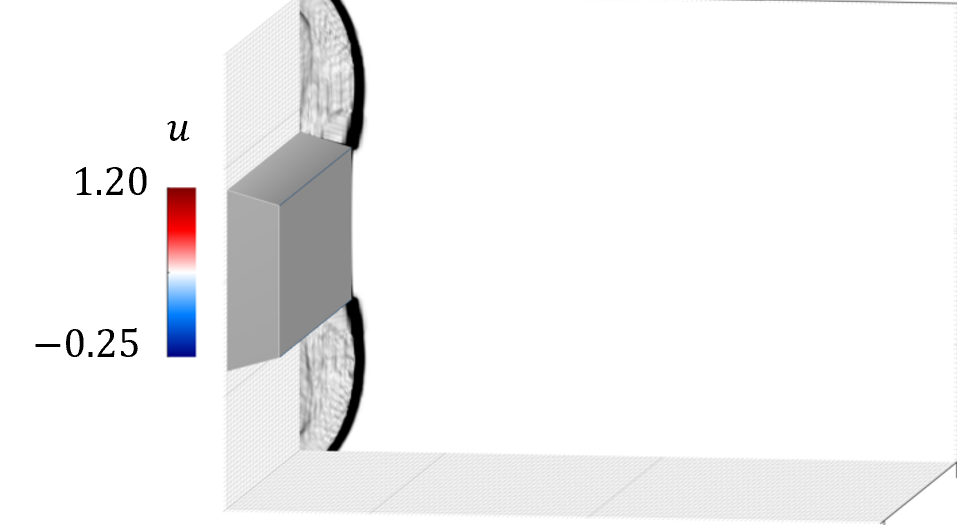}
        \caption{ }
        \label{fig:VortMag3D1}
    \end{subfigure}
    \hfill
    \begin{subfigure}[t]{0.45\textwidth}
        \includegraphics[width=1.0\textwidth]{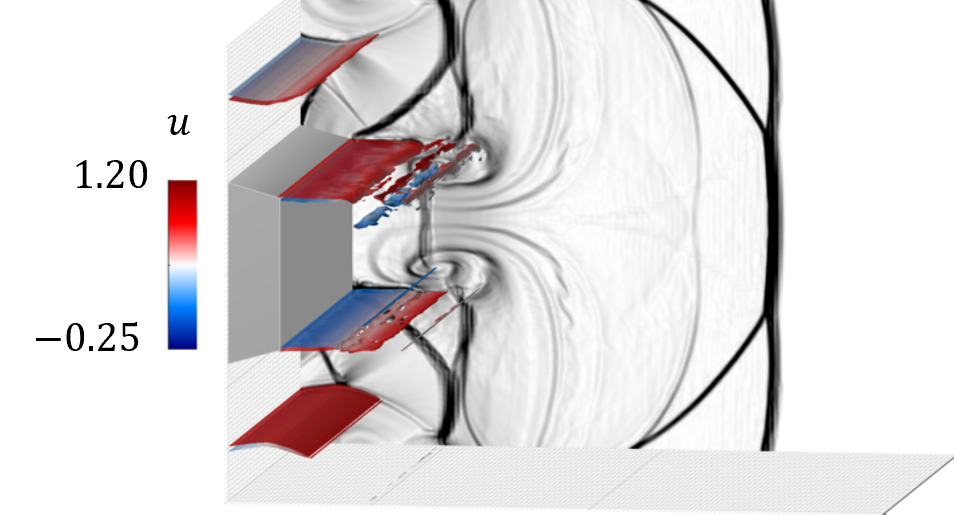}
        \caption{ }
        \label{fig:VortMag3D2}
    \end{subfigure}
    \hfill
    \begin{subfigure}[t]{0.45\textwidth}
        \includegraphics[width=1.0\textwidth]{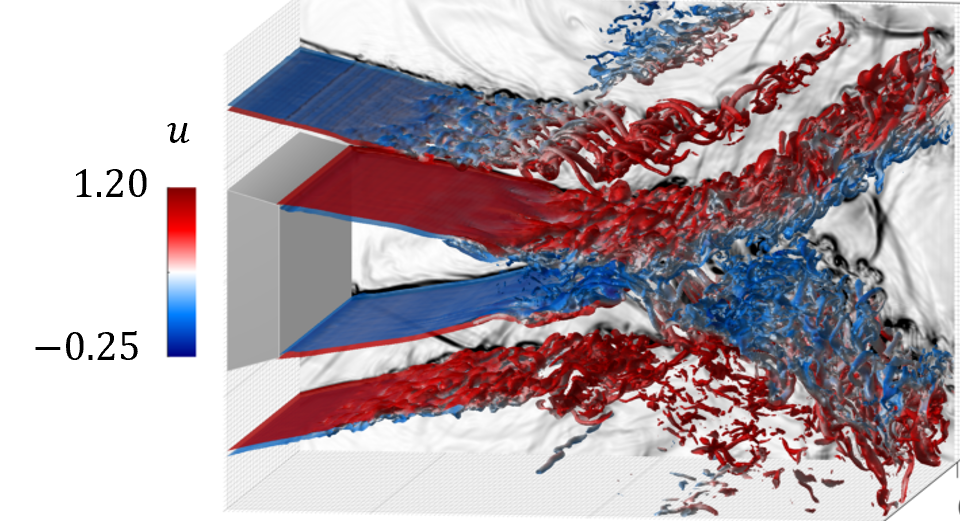}
        \caption{ }
        \label{fig:VortMag3D3}
    \end{subfigure}
    \hfill 
    \begin{subfigure}[t]{0.45\textwidth}
        \includegraphics[width=1.0\textwidth]{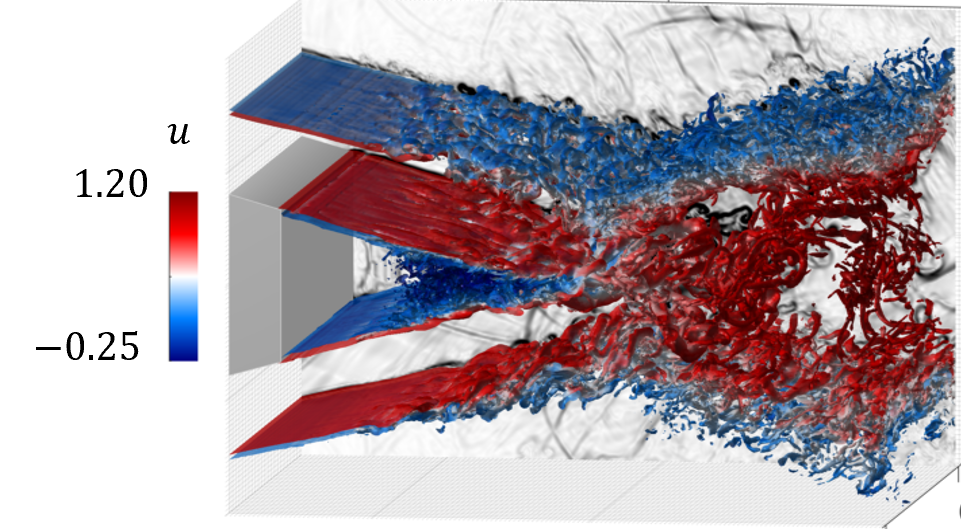}
        \caption{ }
        \label{fig:VortMag3D4}
    \end{subfigure}
    \hfill
    \caption{\rev{Iso-contours of the vorticity magnitude at $\| \omega \| = 12.5$ at times (a) $t = 2.25$, (b) $t = 6.25$, (c) $t = 15$, and (d) $t = 100$.}
    }
    \label{fig:VortMag3D}
\end{figure}

To understand how the three-dimensional flow field is affected by the \Hybrid scheme, we investigate the turbulent statistics extracted over $(7, 0, z)$ and $(1.75, 0, z)$, indicated by the pink lines in \cref{fig:ProbePic} at locations (1) and (2), respectively.
From three-dimensional (\cref{fig:QCriterion3D,fig:VortMag3D}) and two-dimensional (\cref{fig:ProbePic}) visualizations, we expect these regions to  be characterized by turbulence and associated cascade dynamics.

Following a  standard turbulence statistical analysis of compressible Navier-Stokes solutions, we first compute the Favre mean and fluctuations, denoted by a tilde and double prime, respectively,
\begin{equation}
    \tilde{\mathbf{u}} \;=\; \frac{\overline{\rho \mathbf{u}}}{\overline{\rho}}, \qquad \mathbf{u}''(t) \;=\; \mathbf{u}(t) - \tilde{\mathbf{u}},
\label{eq:favrefluct}
\end{equation}
where the overbar identifies a Reynolds average variable. The corresponding density-weighted fluctuations are determined as 
\begin{equation}
    \mathbf{a}(t) \;=\; \sqrt{\rho(t)}\,\mathbf{u}''(t) , 
\label{eq:massweight}
\end{equation}
at time $t = 175$.
\begin{figure}
    \centering
    \includegraphics[width=1.0\linewidth]{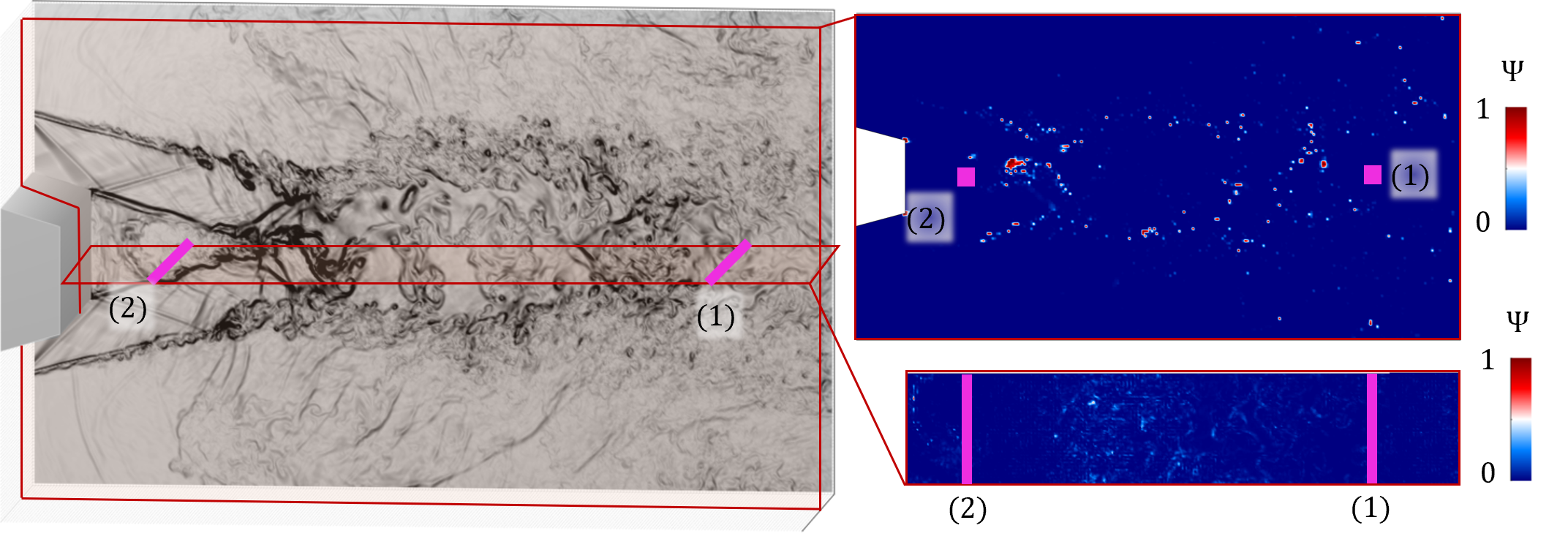}
    \caption{\rev{Iso-contours of $\mu_h$ and contour plots of $\alpha_m$ along the $z=0.5$ and $y=0$ planes at $t = 100$. Profile extraction locations (1) and (2) are pink lines.}
    }
    \label{fig:ProbePic}
\end{figure}
Then, we compute the one-sided power spectral densities of the mass-weighted velocity fluctuations as a function of wavenumber, $\Phi_{ii}(k_3)$, using the Welch method \cite{Welch1967_IEEE}.
Summing the $u$-, $v$-, and $w$-velocity components yields the energy spectrum
\begin{equation}
    \Phi_{tot}(k_3) \;=\; \frac{\Phi_{11}(k_3)+\Phi_{22}(k_3)+\Phi_{33}(k_)}{2}.
\label{eq:totpsd}
\end{equation}
We plot the velocity spectra against the non-dimensional wavenumber in the periodic (z) direction, $k_3 \Delta$, where $\Delta = \overline{(\Delta x + \Delta y  + \Delta z)^{1/3}}$ across profiles (1) and (2) in \cref{fig:VelSpecLoc2} and \cref{fig:VelSpecLoc3}.
The well-known  $-5/3$ slope of the inertial range is recovered for $k_3\Delta \in [10^{-1}, 10^{1}]$.
% demonstrating the inertial range of the turbulence cascade is resolved.
At $k_3 \Delta = 1$, a roll-off is observed in the spectra extracted across both profiles, an indication of significant dissipation of the subgrid scales for $k_3 < \Delta$.
% In other words, the subgrid scales are dissipated.

\begin{figure}[htpb]
    \centering
    \begin{subfigure}[t]{0.33\textwidth}
        \includegraphics[width=\linewidth]{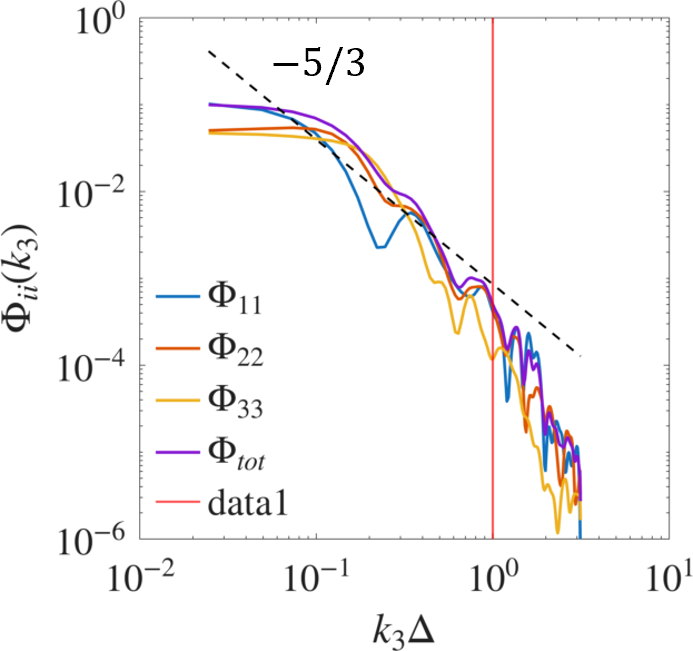}
        \caption{ }
        \label{fig:VelSpecLoc2}
    \end{subfigure}
        \begin{subfigure}[t]{0.33\textwidth}
        \includegraphics[width=\linewidth]{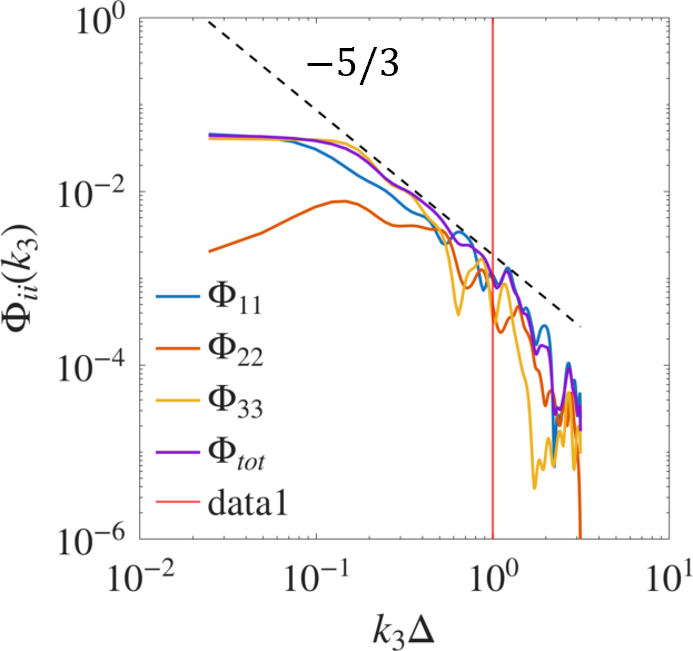}
        \caption{ }
        \label{fig:VelSpecLoc3}
    \end{subfigure}
    \begin{subfigure}[t]{0.33\textwidth}
        \includegraphics[width=\linewidth]{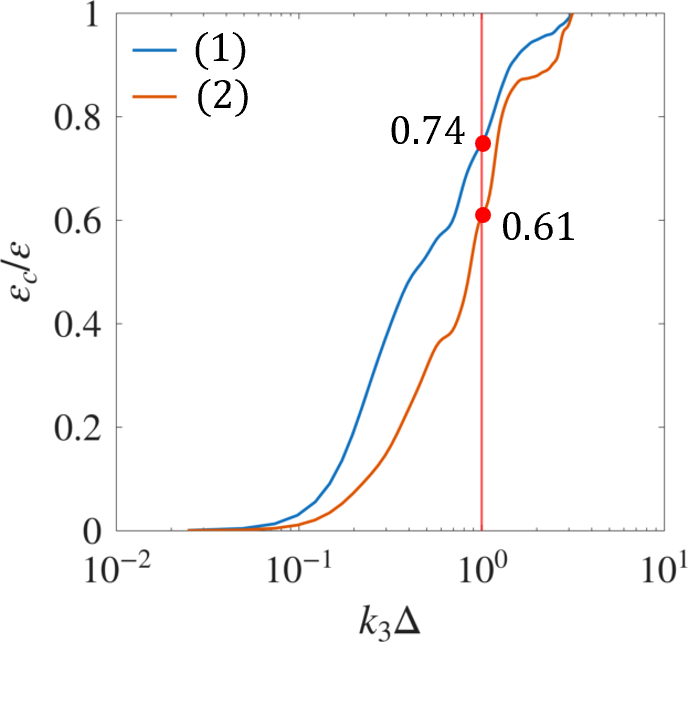}
        \caption{ }
        \label{fig:CumDiss}
    \end{subfigure}
    \caption{\rev{ 1D velocity spectra across profiles (a) 1 and (b) 2, and
    (c) cumulative dissipation. 
    Red dots highlight resolved dissipation fractions at resolution cutoff (red line).
    }
    }
    \label{fig:TurbPlots}
\end{figure}

To determine whether this is a physical or numerical dissipation, we compute the dissipation rate as defined by Pope \cite{Pope},
\begin{equation}
    \varepsilon = 15 \nu \overline{(\frac{\partial w}{\partial z})^2}.
\label{eq:dissDef}
\end{equation}
The $1D$ spectral Parseval identity for the streamwise derivative states
\begin{equation}
    \overline{(\frac{\partial w}{\partial z})^2} = \int_0^\infty k_3^2 \Phi_{33}(k_3)dk_3 \approx \int_0^{k_{max}} k_3^2 \Phi_{33}(k_3)dk_3,
\label{eq:1DSpectThm}
\end{equation}
where $\Phi_{33}$ is the velocity spectrum in the periodic (z) direction and $k_{max}$ is the maximum resolved wavenumber of the extracted velocity spectrum.
Substituting \cref{eq:1DSpectThm} into \cref{eq:dissDef} yields
\begin{equation}
     \varepsilon \approx \int_0^{k_{max}} k_3^2 \Phi_{33}(k_3)dk_3.
\label{eq:dissDef2}
\end{equation}
From \cref{eq:dissDef2}, we compute the Kolmogorov length scale as $\eta = (\frac{\nu^3}{\varepsilon})^{1/4}$ \cite{Pope} and find $\Delta/\eta = 28.6$ and $18.9$ across profiles (1) and (2), respectively.
This clearly demonstrates that the Kolmogorov scales are not resolved and that the energy roll-off in \cref{fig:VelSpecLoc2} and \cref{fig:VelSpecLoc3} is caused by numerical dissipation at the subgrid scales.

However, we do not employ an explicit subgrid-eddy-viscosity SGS model. 
Moreover, from the iso-contours of $\mu_h$ and $\alpha_m$ extracted along the $z=0.5$ and $y=0$ planes in \cref{fig:ProbePic}, we conclude that the no significant numerical dissipation is introduced via the \EV-\FVSE shock-capturing at the profile extraction locations.
Instead, the dissipation of the subgrid scale is implicitly modeled through the minimum global blending coefficient which is set to  $\alpha_{min} = 0.0001$.
The \FVSE scheme  thus contributes everywhere to an implicit diffusion that scales according to the Lax-Friedrich method with the change of solution across a (subcell) element face.
% : $(\Tilde{\mathbf{Q}}_i - \Tilde{\mathbf{Q}}_{i-1})$ or $(\Tilde{\mathbf{Q}}_{i+1} - \Tilde{\mathbf{Q}}_{i})$.
% In smooth solutions this  jump condition scales as $\mathcal{O}(\Delta^{N+1})$, so the dissipation vanishes as $\Delta \to 0$ and $\Tilde{\mathbf{Q}}_i \to \Tilde{\mathbf{Q}}_{i\pm1}$ \cite{Sun2020_arXiv}.
In regions with increasingly uniform and smooth solutions, this dissipation vanishes,  consistent with implicit LES (ILES), in which a vanishing subgrid dissipation is supplied by the numerical scheme's intrinsic dissipation rather than an explicit eddy-viscosity model \cite{Margolin2006_JT, Grinstein}.
Thus, the \Hybrid scheme provides a means to stabilize shocks and implicitly model energy dissipation at the subgrid scales.

To assess the contribution of the \Hybrid to the turbulence dissipation rate, we compute the normalized cumulative dissipation rate as follows:
\begin{equation*}
    \frac{\varepsilon_c(k_3)}{\varepsilon}= \frac{\int_0^{k_3} \kappa^2 \Phi_{33}(\kappa)d\kappa}{\int_0^{k_{max}}k_3^2 \Phi_{33}(k_3)dk}.
\end{equation*}
In \cref{fig:CumDiss}, we see approximately $61\%$ and $74\%$ of the turbulent dissipation is captured by wavenumbers $k_3 \Delta \ge 1$ across profiles (1) and (2), respectively.
In other words, approximately $35\%$ to $40\%$ of turbulent dissipation is implicitly modeled.
These values are consistent with those reported by Komen et al. \cite{Komen2017_JCP}, who report that numerical dissipation accounts for $35\%$ of the viscous dissipation in their ILES of turbulent channel flows
Komen et al. further demonstrate that time-averaged velocity profiles extracted from their ILES are within $4\%$ of time-averaged DNS, and velocity fluctuations and the turbulent kinetic energy budget are within $8\%$.
Hence, the \Hybrid scheme functions as both a shock-capturing scheme and an implicit subgrid viscosity model if $0 < \alpha_{min} << 1$.
% Of course, in the vicinity of discontinuities, i.e., a shock, the penalty term is finite regardless of grid resolution and a local dissipation is introduced that damps the smallest turbulent scales.

}

\FloatBarrier

\section{Conclusions}
\label{sec:Conclusion}

We have developed a kinetic-energy-preserving, discontinuos Galerkin spectral element method (\DGSEM), Navier-Stokes solver that captures shocks with a hybridization of the  entropy viscosity (\EV) and the finite-volume subcell element (\FVSE) method, for high-fidelity computations of the unsteady aerospike nozzle flow.
The \FVSE scheme acts as a numerical "parachute" that is deployed only where gradients cannot be smoothed by \EV without incurring prohibitively small time steps. 
The efficacy of this solver is demonstrated in three aerospike test cases: two-dimensional perfectly expanded flow, two-dimensional underexpanded flow, \rev{and three-dimensional perfectly expanded flow.}

\rev{
The convergence rate of the \Hybrid scheme approximating the solution with an $N^{th}$ order polynomial is $N+1$ in the limit of smooth flows, which is consistent with theoretical convergence rate of smooth \DGSEM solutions. 
The computational overhead of the \Hybrid scheme is, at most,  $\approx 25 \%$ greater than the \DGSEM with the \EV scheme. 
Overhead may be reduced via an adaptive \FVSE activation algorithm.
Computations using \EV  and a spectral filter (EVF scheme) are unstable for computational times greater than $23$ convective time units; 
the spectral filter does not enforce monotonicity across shocks and numerical oscillations can violate pressure and density positivity.
Applying a constant \FVSE blending to all elements suppresses wave motions, shear instabilities, and vortices that are otherwise present, resulting in a solution qualitatively comparable to that of a RANS model.
\EV-based adaptive blending models introduce numerical dissipation locally in shocked elements, preserving boundedness without over-damping the flow field.
\FVSE blending is indirectly attenuated through the density-scaled \EV tuning parameter, $C_m$.
As a result, the \EV-\FVSE scheme is intrinsically more dissipative in underexpanded, i.e., high Mach number, aerospike nozzle flows, and pressure positivity is maintained even in the hypersonic (Mach $6$) regime without manual tuning of the $C_m$ coefficient.
}

\rev{
% Monotone solutions are recovered over shocks with upstream Mach numbers up to $6$.
Numerical oscillations observed in non-stabilized, shock-adjacent elements are mitigated by increasing the minimum \FVSE blending coefficient to $\alpha_{min} = 0.0001$.
The global numerical dissipation introduced by the $\alpha_{min} > 0$ blending implicitly models the turbulent dissipation of the unresolved subgrid scales, evidenced by a roll-off at $k_1 \Delta = 1$ in the velocity spectrum.
A $-5/3$ slope over $k_1 \Delta \in[0.25, 1]$ demonstrates turbulence is resolved down to the inertial scales.

Preliminary analysis of the flow shows several instabilities that are not currently reported in literature. We reserve a detailed discussion of the flow physics for future work. 
}

\section*{Acknowledgments}
This work was supported by the Department of Defense Air Force Research Laboratory by grant FA930025C6002.
Additional funding was provided by the San Diego State University Graduate Fellowship and the American Institute of Aeronautics and Astronautics (AIAA) Reuben H. Fleet Scholarship.
The computational time for this project was made available through the Department of Defense High Performance Computing Modernization Program (HPCMP).

\clearpage
\bibliography{references}

\end{document}